\documentclass[aps,prc,reprint,superscriptaddress]{revtex4-2}

\usepackage{graphicx}

\begin{document}

\title{Total cross sections for the reactions\\$^{10,11,12}$Be~+~$^{28}$Si and $^{14}$B~+~$^{28}$Si}

\author{Yu.~G.~Sobolev}
\affiliation{Flerov Laboratory of Nuclear Reactions, Joint Institute for Nuclear Research, Dubna 141980, Russia}

\author{S.~S.~Stukalov}
\affiliation{Flerov Laboratory of Nuclear Reactions, Joint Institute for Nuclear Research, Dubna 141980, Russia}
\affiliation{Voronezh State University, Voronezh, 394018, Russia}

\author{Yu.~E.~Penionzhkevich}
\affiliation{Flerov Laboratory of Nuclear Reactions, Joint Institute for Nuclear Research, Dubna 141980, Russia}
\affiliation{Department of Experimental Methods in Nuclear Physics, National Research Nuclear University, Moscow 115409, Russia}

\author{M.~A.~Naumenko}
\email{anaumenko@jinr.ru}
\affiliation{Flerov Laboratory of Nuclear Reactions, Joint Institute for Nuclear Research, Dubna 141980, Russia}

\author{V.~V.~Samarin}
\affiliation{Flerov Laboratory of Nuclear Reactions, Joint Institute for Nuclear Research, Dubna 141980, Russia}
\affiliation{Department of Nuclear Physics, Dubna State University, Dubna 141982, Russia}

\date{\today}

\begin{abstract}
In this paper, the results of measurements of the total cross sections for the reactions $^{10,11,12}$Be~+~$^{28}$Si and $^{14}$B~+~$^{28}$Si in the beam energy range $13A$--$47A$~MeV are presented. The experimental cross sections were obtained by detection of the gamma quanta and neutrons accompanying the interaction of the isotopes of Be and B with $^{28}$Si. It was found that the cross sections for $^{11,12}$Be are similar, but significantly exceed those for $^{10}$Be. A significant increase in the cross sections for $^{12}$Be with decreasing energy is observed in the entire measured energy range. A theoretical explanation of the obtained experimental data is given based on the microscopic model of deformed nuclei and the numerical solution of the time-dependent Schr\"{o}dinger equation for the outer weakly bound neutrons of the projectile nuclei. The calculated total reaction cross sections are in good agreement with the experimental data.
\end{abstract}

\maketitle

\section{\label{sec:sec01} Introduction}
This work is aimed at measurements and theoretical calculations of the total cross sections for the reactions $^{10,11,12}$Be~+~$^{28}$Si and $^{14}$B~+~$^{28}$Si in the previously unexplored energy range $13A$--$47A$~MeV.

Among light nuclei, clustering is especially pronounced in the isotopes of beryllium. For example, their moment of inertia turned out to be very large, which is consistent with their $2\alpha$-cluster structure characterized by a large deformation (e.g.,~\cite{penionzhkevich_reactions_2011}). In addition, small values of the alpha-particle separation energies may be an indication the alpha cluster structure of the Be isotopes. Calculations of the $^{A}$Be isotopes with $A\ge8$ also yield a dumbbell-shaped structure with the pronounced $2\alpha$-clustering~\cite{Freer_2007, Zhao_2021}. The $^{10}$Be and $^{10}$B nuclei may be represented as composed of two $\alpha$-clusters and two outer (valence) weakly bound nucleons~\cite{Freer_2007, Zhao_2021, Samarin_2022A, Samarin_2020}. These nuclei may be regarded as simple two-center nuclear molecules~\cite{Oertzen_1970, Okabe, Scharnweber}. Thus, the $\alpha$-cluster model explains the large deformation of the $^{10}$Be and $^{10}$B nuclei.

Some properties of the $^{9-12}$Be and $^{10-14}$B nuclei (neutron, proton, and alpha-particle separation energies, rms charge radii, and quadrupole deformations) are given in Table~\ref{tab:tab01}. It can be seen that all listed nuclei have similar charge radii, but despite an additional proton in the B isotopes, their charge radii are somewhat smaller than those for the Be isotopes, which may be explained by their smaller quadrupole deformations. The $^{11}$Be nucleus has the least-bound neutron among all nuclei listed in Table~\ref{tab:tab01} and can be represented as a system of a deformed nuclear core, similar in properties to the $^{10}$Be nucleus, and a weakly bound neutron. The sequence of the Be and B isotopes allows one to trace the change in their properties with the filling of nucleon shells.
\begin{table}[htbp]
\caption{\label{tab:tab01} Some properties of the $^{9-12}$Be and $^{10-14}$B nuclei: spin-parity $J^\pi$, neutron ${S_{1n}}$, proton ${S_{1p}}$, and alpha-particle ${S_{1\alpha }}$ separation energies, rms charge radius ${\left\langle {r_{{\rm{ch}}}^2} \right\rangle}^{1/2}$ (taken from~\cite{zagrebaev_nrv_1999}), and quadrupole deformation parameter $\beta_2$ (taken from~\cite{cdfe}; for the deformations marked by *, sign is unknown).}
\begin{ruledtabular}
\begin{tabular}{ccccccc}
Nucleus &$J^{\pi}$ & ${S_{1n}}$ & ${S_{1p}}$ & ${S_{1\alpha }}$ & ${\left\langle {r_{{\rm{ch}}}^2} \right\rangle ^{{1 \mathord{\left/ {\vphantom {1 2}} \right. \kern-\nulldelimiterspace} 2}}}$ & $\beta_2$\\
& & (MeV) & (MeV) & (MeV) & (fm) & \\
\colrule
$^{9}$Be & ${3/2}^-$ & 1.665 & 16.886 & 2.462 & 2.518~$\pm$~0.0119 & +0.89\\
$^{10}$Be & $0^+$ & 6.812 & 19.636 & 7.409 & 2.361~$\pm$~0.017 & 1.13*\\
$^{11}$Be & ${1/2}^+$ & 0.502 & 20.164 & 8.320 & 2.466~$\pm$~0.015 & \\
$^{12}$Be & $0^+$ & 3.171 & 22.940 & 8.957 & 2.503~$\pm$~0.015 & \\
$^{10}$B & $3^+$ & 8.437 & 6.587 & 4.461 & 2.4277~$\pm$~0.0499 & +0.52\\
$^{11}$B & ${3/2}^-$ & 11.454 & 11.229 & 8.664 & 2.406~$\pm$~0.0294 & +0.498\\
$^{12}$B & $1^+$ & 3.370 & 14.097 & 10.001 & & 0.157*\\
$^{13}$B & ${3/2}^-$ & 4.879 & 15.804 & 10.817 & & 0.418*\\
$^{14}$B & $2^-$ & 0.970 & 17.284 & 11.813 & & 0.228*\\
\end{tabular}
\end{ruledtabular}
\end{table}

No more than two neutrons can be present on each energy level of a deformed nucleus, therefore with an increase in the number of neutrons from five in $^{9}$Be to nine in $^{14}$B, the energy and density distribution of outer neutrons will be very different, which can manifest itself in the total reaction cross sections. Thus, measurements of the total reaction cross sections for the $2\alpha$-core $^{10}$Be nucleus and neutron-rich $^{11,12}$Be and $^{14}$B nuclei may provide information on their structures. One may also obtain information on such properties as effective matter radii, because the values of the total reaction cross sections are mainly determined not by the charge distribution, but by the matter distribution, which for the nuclei listed in Table~\ref{tab:tab01}, is mainly determined by the distribution of neutrons.

In the analysis of elastic scattering data within the optical model, the parameters of the optical potential are usually determined by fitting of the results of calculations to the experimental data. However, it is well known that there are large uncertainties in the optical potential. These uncertainties can be significantly reduced by the simultaneous analysis of the data on elastic scattering and total reaction cross sections. In this work, this approach allowed us to determine the effective matter radii of the Be isotopes. In addition, for the $^{11,12}$Be and $^{14}$B nuclei, we used the optical model to determine the cross sections of the reaction channels involving their core $^{10}$Be.

For a theoretical description of the total reaction cross sections within the framework of a microscopic model, we used the results of calculations in the shell model of the deformed nucleus. For shell-model calculations, it is important that they lead to the correct spin-parity $J^\pi$ and nucleon separation energies for the ground states of nuclei. For example, in the $^{9,11}$Be nuclei, the quantum numbers of the upper partially filled neutron shells are $3/2^-$ and $1/2^+$, respectively. In the $^{11,13}$B nuclei, the quantum number of the upper partially filled neutron shell is $3/2^-$. The quantum numbers of the upper partially filled neutron shells for the ground states of $^{10,12,14}$B can be determined from the values of the spins of the nuclei taking into account the momentum addition rule. The values of the neutron separation energies were used to determine the depths of the potential wells for the studied nuclei.

The calculations of the total cross sections for the reactions with the $^{11,12}$Be and $^{14}$B nuclei were performed in a similar way as for the $^{11}$Li nucleus~\cite{Penionzhkevich_2019, Azhibekov_2022}. The calculations were based on the assumption that the total cross sections for reactions involving weakly bound nuclei can be represented in the approximate form:
\begin{equation}
{\sigma _{\rm{R}}}\approx {\sigma _{\rm{R}}}^{(\rm{core})} + {\sigma _{ - xn}},
\label{eq:eq01}
\end{equation}
as shown experimentally in ~\cite{tanihata_revelation_1992, warner_total_1996} for the $^{6}$He and $^{11}$Li nuclei with $x=2$. The contribution of the neutron removal channels ${\sigma _{ - xn}}$ was determined in the semiclassical model based on numerical solution of the time-dependent Schr\"{o}dinger equation~(TDSE) for an outer neutron of a projectile nucleus (e.g.,~\cite{samarin_2015}) with the initial conditions obtained from the calculations within the shell model of the deformed nucleus.

Section~\ref{sec:sec02} provides a brief overview of the methods for direct measurement of the total cross sections in reactions with radioactive nuclei produced at fragment separators. Section~\ref{sec:sec03} describes the experimental setup and the experimental procedure. Section~\ref{sec:sec04} is devoted to determining the detection efficiency of gamma quanta with different multiplicities on the used multi-detector spectrometer. Section~\ref{sec:sec05} describes the procedure of processing of the obtained experimental data for the reactions with the $^{10,11,12}$Be and $^{14}$B nuclei taking into account the number of triggered spectrometer detectors. Section~\ref{sec:sec06} presents the measurement results along with their discussion. 
\section{\label{sec:sec02} Experimental methods for direct measurement of total reaction cross sections}
\begin{figure}[htbp]
\includegraphics[width=7.6cm]{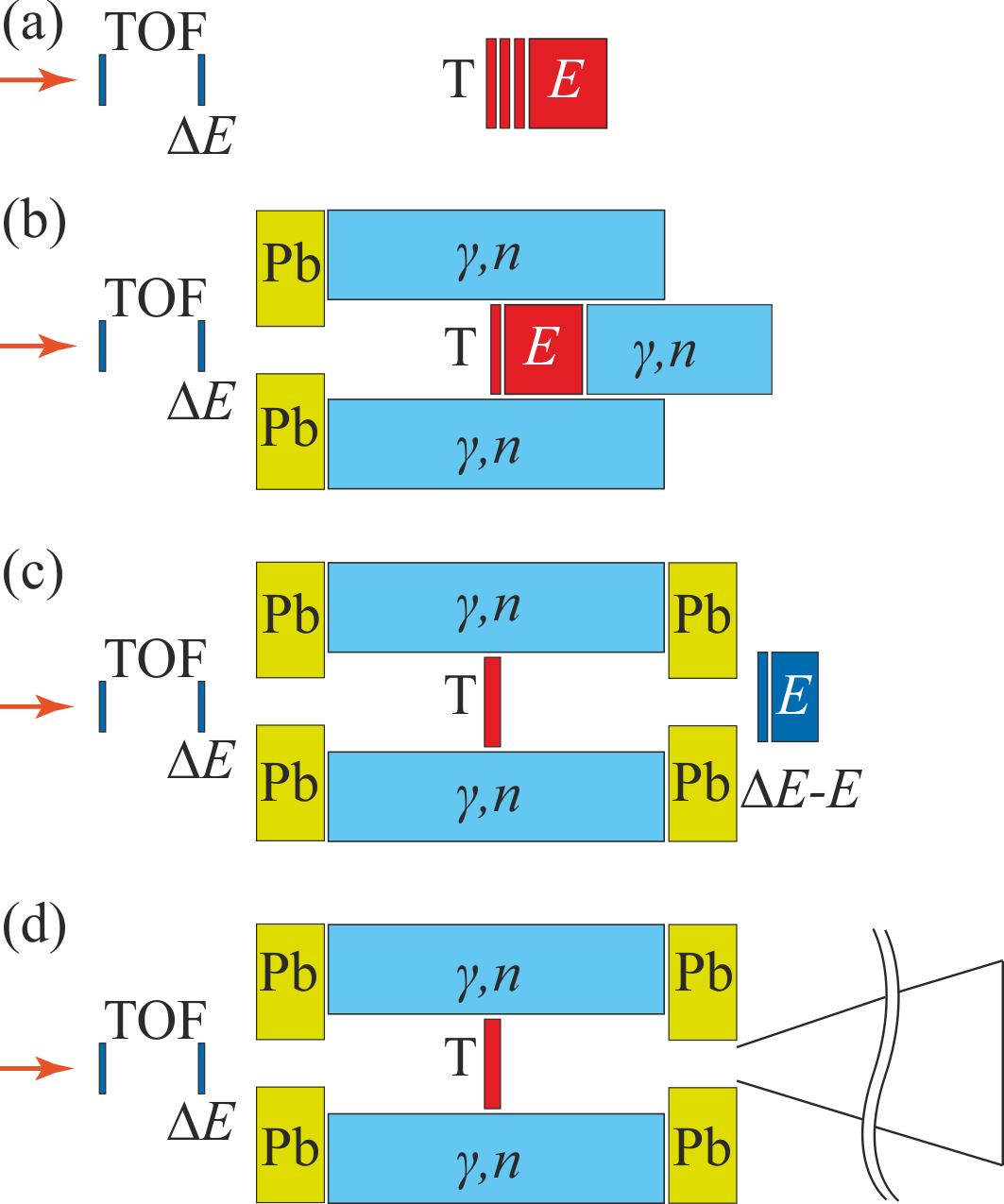}
\caption{\label{fig:fg01} Typical detector configurations for measuring total reaction cross sections using transmission telescopes and 4$\pi$ scintillation spectrometers: $\gamma, n$ denotes the scintillation detectors of the spectrometer; TOF and $\Delta E$ are the detectors for identifying particles in front of the target T by the time-of-flight technique; $E$ and $\Delta E$-$E$ are the detectors of the telescopes installed as a target (a) and behind the target (b), (c); Pb denotes the elements of protection of the spectrometer.}
\end{figure}

Most experiments for measurement of total reaction cross sections ${\sigma _{\rm{R}}}$ with exotic nuclei were performed by direct methods using secondary beams. The secondary beams are typically formed by fragment separators and are a cocktail of the products of nuclear reactions between the particles of a primary beam and the nuclei of a production target. The intensity of the secondary beams of radioactive nuclei usually does not exceed 10$^{3}$~pps~\cite{rodin_status_2003}, which makes it possible to register every event, when a beam particle hits the target. Typical detector configurations for measuring total reaction cross sections using transmission telescopes and 4$\pi$ scintillation spectrometers are shown in Fig.~\ref{fig:fg01}.

The system for identification of secondary beam particles consists of thin detectors located along the beam axis in front of the target. The particles are identified by the $\Delta E$-TOF method: the ionization losses $\Delta E$ of the particles in the matter of a thin Si detector installed in front of the target and the time of flight TOF on a certain distance are measured. Typically, TOF is measured by fast scintillation detectors installed at the beginning point and at the end point of the distance.

In experiments, certain values of the magnetic rigidity $B \rho$ are set for the dipole magnets of the fragment separator, where $B$ is the magnetic field induction and $\rho$ is the radius of the circular trajectory of the particles. For a given value of $B \rho$, several groups of particles with similar values of ${{vA} \mathord{\left/ {\vphantom {{vA} Z}} \right. \kern-\nulldelimiterspace} Z} \sim B\rho $ will be present in the secondary beam at the exit of the separator, where $v$ is the particle velocity, $A$ and $Z$ are its mass and charge numbers, respectively. Isotopes will fill different regions in the $\Delta E$-TOF matrix according to $A$ and $Z$, which makes it possible to identify the particles, which hit the target, and to determine their number $I_0$.

The system for identification of the beam particles, which hit the target is a common and necessary part of all experimental methods for direct measurement of the total reaction cross sections ${\sigma _{\rm{R}}}$. The differences between these methods lies in the ways of detection of the reaction products and/or beam particles that passed through the target without interaction. In this respect, experiments can be divided into two groups.

The first group includes the experiments carried out by the transmission~\cite{x} or attenuation~\cite{xx} method. This method, developed for beams of stable nuclei, has been modernized and successfully used up to now for beams of radioactive nuclei~\cite{xxx, xxxx}. In the performed experiments, the transmission telescopes $\Delta E$-…-$\Delta E$-$E$ [Fig.~\ref{fig:fg01}(a)] were mainly used for detection of the beam particles that passed through a thin target without interaction with target nuclei and to measure the number of such particles $I$.

In the approximation of independence of the total reaction cross section ${\sigma _{\rm{R}}}$ from energy, it was determined from the expressions:
\begin{equation}
I = {I_0}\exp \left( { - \kappa {\sigma _{\rm{R}}}x} \right) = {I_0}\exp \left( { - n{\sigma _{\rm{R}}}} \right),
\label{eq:eq02}
\end{equation}
\begin{equation}
\sigma _{\rm{R}} = - \frac{1}{n}\ln \frac{I}{{I_0 }},
\label{eq:eq03}
\end{equation}
where $\kappa = {N_{\rm{A}}}{\rho \mathord{\left/ {\vphantom {\rho m}} \right. \kern-\nulldelimiterspace} m}$ is the concentration of the target nuclei, ${N_{\rm{A}}}$ is the Avogadro constant, $\rho$ is the density of the target matter, $m$ is the molar mass of the target matter, $n = \kappa x$ is the number of the target nuclei per unit area, and $x$ is the target thickness.

This group of experiments has several difficulties associated with the analysis of obtained energy spectra. In particular, it is difficult to distinguish inelastically scattered beam particles from the particles that passed through the target without interaction and the nuclear reactions that occurred in the target -- from the nuclear reactions in the $E$-detector. These problems are exacerbated by the energy spread of secondary beams of fragment separators. These difficulties have been partially eliminated in the second group of experiments for direct measurement of total reaction cross sections.

The second group includes experiments, in which scintillation spectrometers detected prompt gamma quanta and neutrons emitted during a reaction in a solid angle close to the total solid angle 4$\pi$. This way, the number of reaction events $\Delta I$ was measured and the reaction probability ${P_{\rm{R}}} = {{\Delta I} \mathord{\left/ {\vphantom {{\Delta I} {{I_0}}}} \right. \kern-\nulldelimiterspace} {{I_0}}}$ was determined. For a thin target of thickness $x$,
\begin{equation}
\Delta I = {I_0}\left[ {1 - \exp \left( { - \kappa {\sigma _{\rm{R}}}x} \right)} \right] = {I_0}\left[ {1 - \exp \left( { - n{\sigma _{\rm{R}}}} \right)} \right]
\label{eq:eq04}
\end{equation}
and the total reaction cross section is
\begin{equation}
{\sigma _{\rm{R}}} = - \frac{1}{{\kappa x}}\ln \left( {1 - {P_{\rm{R}}}} \right) = - \frac{1}{n}\ln \left( {1 - {P_{\rm{R}}}} \right).
\label{eq:eq05}
\end{equation}

Typical detector configurations for measuring ${\sigma _{\rm{R}}}$ with 4$\pi$ scintillation spectrometers consisting of several detectors are shown in Fig.~\ref{fig:fg01}(b)--(d). The efficiency of detection of gamma quanta by these spectrometers was determined using calibrated sources.

The configuration of scintillation detectors with a thick target (silicon detector telescope) in Fig.~\ref{fig:fg01}(b) was used to measure the mean energy-integrated reaction cross section ${\sigma _{\rm{R}}}$ on Si~\cite{Mittig, Villari, Licot, Khouaja}. In the case of a thick target, the particles stopped completely at the range ${R=R_{\max }}$, and the number $\Delta I$ of reaction events was determined by the expression:
\begin{eqnarray}
&&\ln \frac{{{I_0} - \Delta I}}{{{I_0}}} = \ln (1 - {P_{\rm{R}}}) = - \kappa \int\limits_0^{{R_{\max }}} {\sigma [E(R)]dR}\nonumber\\
&&= - \kappa \int\limits_{{E_{\min }}}^{{E_{\max }}} {\sigma [E(R)]\frac{{dR}}{{dE}}dE}.
\label{eq:eq06}
\end{eqnarray}
In~\cite{Villari, Khouaja}, the mean energy-integrated reaction cross section was determined by analogy with formula~(\ref{eq:eq05}) for a thin target, by the expression:
\begin{eqnarray}
&&{\bar \sigma _{\rm{R}}} = - \frac{1}{{\kappa {R_{\max }}}}\ln \left( {1 - {P_{\rm{R}}}} \right)\nonumber\\
&&= - \frac{m}{{{\rho _{{\rm{Si}}}}{N_{\rm{A}}}{R_{\max }}}}\ln \left( {1 - {P_{\rm{R}}}} \right),
\label{eq:eq07}
\end{eqnarray}
where ${\rho _{{\rm{Si}}}}$ is the density of silicon. The upper limit ${E_{{\rm{max}}}}$ of the energy range was determined by the energy of the beam particles, the lower limit ${E_{{\rm{min}}}}$ -- by the reaction threshold. The $\Delta E$-$E$ detectors of the telescope were used as a target. The scintillation detectors of the spectrometer registered prompt gamma quanta and neutrons from the reactions occurring in the matter of the $\Delta E$-$E$ telescope until the particle stopped. To determine the number of events $\Delta I$, the energy spectrum of particles was accumulated in the $E$-detector under the condition that the signals from the $E$-detector coincided with those from the detectors of the 4$\pi$ spectrometer. In~\cite{Mittig}, the authors used the detection efficiency of the 4$\pi$ spectrometer equal to 84\%; it was determined taking into account the solid angle of the spectrometer and the measured multiplicity and exceeded the detection efficiency of one photon which was equal to 70\%.

Significant disadvantages of the scheme of experiments in Fig.~\ref{fig:fg01}(b) are the need of using detectors (Si or Ge) as targets and the need of using the targets, whose thickness must exceed the maximum range ${R_{\max }}$ of particles. The inevitable averaging of the total reaction cross sections over a wide energy range by formula~(\ref{eq:eq07}) is also a disadvantage of the scheme in Fig.~\ref{fig:fg01}(b) with a thick target.

In the configurations with a thin target shown in Fig.~\ref{fig:fg01}(c),(d), the $\Delta E$-$E$ detectors of the telescope are located behind the spectrometer [Fig.~\ref{fig:fg01}(c)] or completely removed from the setup [Fig.~\ref{fig:fg01}(d)]. Such configurations make it possible to measure the energy dependence of the total reaction cross sections ${\sigma _{\rm{R}} (E)}$ with targets made of various matters.

In the configuration of Fig.~\ref{fig:fg01}(c), the close location of the $\Delta E$-$E$ detectors of the telescope to the sensitive zone of the spectrometer is a source of an additional background caused by nuclear reactions in the matter of the telescope~\cite{z, zz}.

The solution of the problem of the background radiation caused by the reactions in the $\Delta E$-$E$ telescope is implemented in the configuration shown in Fig.~\ref{fig:fg01}(d) where the $\Delta E$-$E$ telescope is absent. A cone-shaped scattering chamber is used to prevent the scattered particles from hitting its walls near the spectrometer and thus reduce background gamma quanta and neutrons. Using this configuration, we measured the total cross sections ${\sigma _{\rm{R}} (E)}$ for the reactions $^{6,8}$He, $^{8,9,11}$Li~+~$^{28}$Si, $^{59}$Co, $^{181}$Ta~\cite{Penionzhkevich_2019, Sobolev_2020, www} and $^{10,11,12}$Be, $^{14}$B~+~$^{28}$Si (this work). In this detector configuration, it is not possible to detect gamma quanta and neutrons at small forward angles. However, in the process of breakup of nuclei with low neutron separation energy, a significant number of neutrons are emitted into the forward angles. For example, differential cross sections for neutron emission in the reactions on Au, Ni, and Be targets for $^{11}$Li at the beam energy of 29$A$~MeV measured in~\cite{anne_observation_1990} and of $^{11}$Be at the beam energy of 41$A$~MeV measured in~\cite{anne_observation_1993, anne_observation_1994} were strongly anisotropic with a maximum yield in a narrow range of forward angles. To solve this problem, in this work, we used not a constant value of the spectrometer detection efficiency, but a more accurate approach taking into account the distribution of the number of triggered spectrometer detectors on the multiplicity of gamma quanta and neutrons and an additional correction~\cite{Sobolev_2020, www}.

\section{\label{sec:sec03} Experiment}
\begin{figure}[htbp]
\includegraphics[width=8.6cm]{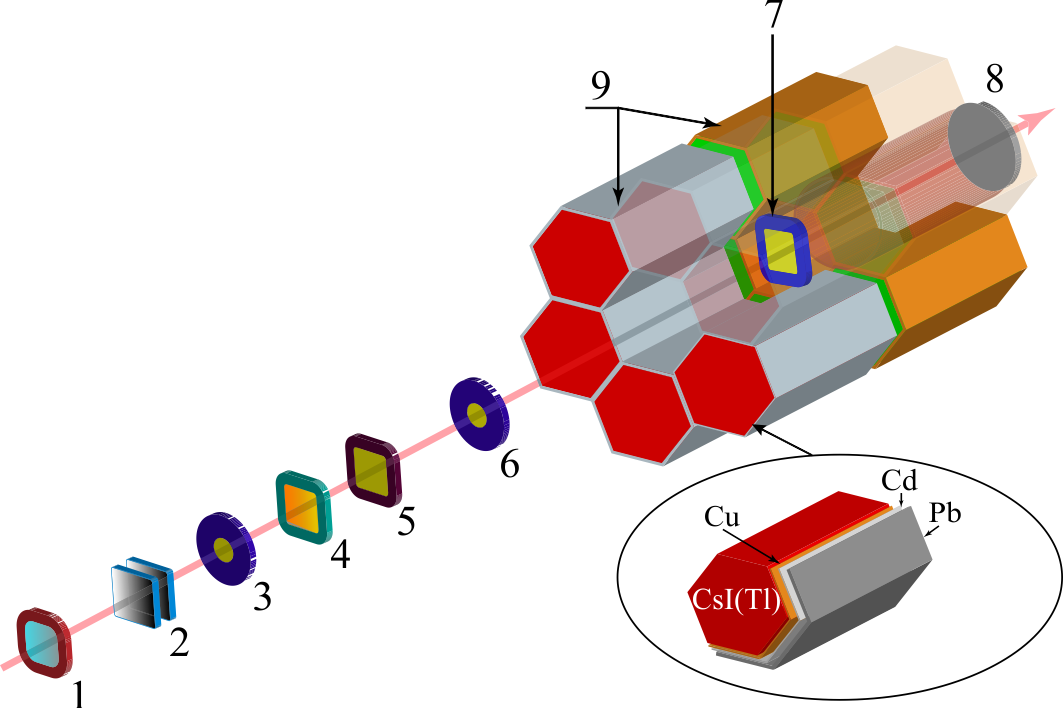}
\caption{\label{fig:fg02} Scheme of the experimental setup: (1) plastic scintillation detector $\Delta {E_{\rm{TOF2}}}$; (2) polyethylene plates; (3) scintillation detector AC1 (PMT is not shown); (4) $\Delta E$ Si strip detector; (5) start $\Delta E_0$ Si detector; (6) scintillation detector AC2 (PMT is not shown); (7) $\Delta {E_{\rm{T}}}$ Si target detector; (8) beam exit window; (9) CsI(Tl) scintillation detectors with~PMTs.}
\end{figure}

The experiments on measuring the energy dependence of the total cross sections for the reactions $^{10,11,12}$Be, $^{14}$B~+~$^{28}$Si were done by a direct method using a CsI(Tl) gamma spectrometer in the configuration of Fig.~\ref{fig:fg01}(d), as mentioned above. The experimental setup and the method for measuring total reaction cross sections by the detection of prompt $\gamma, n$ radiation are described in~\cite{sobolev_experimental_2017, Penionzhkevich_2019}. The experiments were carried out at the U-400M accelerator of the Laboratory of Nuclear Reactions, JINR, on the channel of the achromatic fragment separator ACCULINNA~\cite{rodin_status_2003}. The primary $^{15}$N beam with an energy of 49.7$A$~MeV was focused on the producing $^{9}$Be target. The secondary beam of fragmentation reaction products was formed and purified by the magnetic system of the fragment separator. At the exit of the last dipole magnet of the separator, the beam entered the straight section of the ion beam line, where the system for measuring the time of flight ${T_{{\rm{TOF}}}}$ consisting of two scintillation detectors $\Delta {E_{{\rm{TOF1}}}}$, $\Delta {E_{{\rm{TOF2}}}}$ was located. The scheme of the setup is shown in Fig.~\ref{fig:fg02}.

The energy of the secondary beam particles was varied without significant decrease of intensity by the magnetic system of the fragment separator --- in the energy range of 33.5$A$--47$A$~MeV and by a set of polyethylene plates of various thicknesses installed behind the $\Delta {E_{{\rm{TOF2}}}}$ detector --- in the energy range of 13$A$--34$A$~MeV.

The beam was focused on a position-sensitive $\Delta E$ Si strip detector with a thickness of 300~$\mu$m located in the focal plane of the fragment separator. This detector was used for preliminary tuning of the beam parameters (intensity, profile, and isotopic composition).

The scintillation detectors AC1 and AC2 were used as active collimators; they were installed in such a way that the trajectories of particles passed through the central region of the target detector $\Delta {E_{\rm{T}}}$ (natural Si, 790-$\mu$m thick) and exit window without touching their holders. The $\Delta {E_0}$ Si detector (243-$\mu$m thick) was used to generate the start signal for the data acquisition system for every event of entering of the beam particles into the $\Delta {E_0}$-detector. The ionization losses $\Delta {E_0}$ were used for identification of the beam particles by the $\Delta {E}$-TOF method.

The target detector $\Delta {E_{\rm{T}}}$ was installed on the beam axis in the reaction chamber which was a thin-walled vacuum cylinder of stainless steel. On the outside, the cylinder was surrounded by the gamma spectrometer of twelve CsI(Tl) scintillators which were prisms with a height of 15~cm and a base in the form of a regular hexagon. From one end, each scintillator was optically connected to a photomultiplier (PMT). The side surfaces of the prisms were covered by a reflective coating and thin light-protection layers. Three outer sides of each prism were covered by a protective layer consisting of three Cd, Pb, and Cu plates (each 1-mm thick) for protection from the low-energy background gamma quanta and neutrons. The minimum distance from the center of the target to the surface of the scintillators was $\sim$~4.4~cm, so that the minimum angle ${\theta _{\min }}$ between the axis of the setup and the direction of emission of a gamma quantum or a neutron from the center of the target with the entry into the volume of the scintillators was ${\theta _{\min }=16^{\circ}}$ and the solid angle covered by the scintillators was $\Omega = 4\pi {\eta _0}$, where ${\eta _0} \approx \cos \left( {{\theta _{\min }}} \right)=$~0.96 is the geometric efficiency. Other detectors were located outside the sensitive zone of the gamma spectrometer, which provided a minimum level of background triggering for the gamma detectors. The spectrometer with the reaction chamber was located inside a lead cube with a wall thickness of 5~cm, the outer sides of which were covered by plates of 10-cm-thick boron-containing polyethylene.

The events of nuclear reactions are accompanied by emission of gamma quanta and neutrons which are characterized by the multiplicity $M$ and emission anisotropy. The anisotropy of neutron emission is especially pronounced in reactions with weakly bound nuclei~\cite{anne_observation_1990, anne_observation_1993, anne_observation_1994}. The detection efficiency of the spectrometer $\varepsilon$ depends on the multiplicity $M$ of registered particles -- gamma quanta and neutrons. Gamma quanta and neutrons could not be distinguished by the CsI(Tl) scintillation detectors of the spectrometer because of the insufficient time resolution at the flight distance of $\sim$~4.4~cm. An analysis of the energy dependence of gamma quanta detection efficiency was done in~\cite{Sivacek2020}. Based on the results of~\cite{NIM_A274_1989}, we estimated the efficiency of neutron detection as 20\%. The procedure and results of measuring the detection efficiency of registration for gamma quanta are presented in Section~\ref{sec:sec04}.

The measurement results (raw data) is the lower boundary ${\tilde \sigma_{\rm{R}}}$ of the total reaction cross section ${\sigma _{\rm{R}}} = {{{{\tilde \sigma }_{\rm{R}}}} \mathord{\left/ {\vphantom {{{{\tilde \sigma }_{\rm{R}}}} \eta }} \right. \kern-\nulldelimiterspace} \eta }$, where ${\eta}$ is an empirical correction that takes into account not only the anisotropy of neutron emission~\cite{anne_observation_1990, anne_observation_1993, anne_observation_1994}, but also the contribution of such reaction channels, as elastic breakup channels which are not accompanied by gamma radiation~\cite{star1}. Their contribution to the total cross section ${\sigma _{\rm{R}}}$ for reactions with weakly bound neutron-rich nuclei can be several hundred millibarns~\cite{warner_2001}. To estimate the empirical correction ${\eta}$, the lower boundaries ${ \tilde \sigma _{\rm{R}}}$ of the total reaction cross sections for the $^{10,11,12}$Be and $^{14}$B nuclei were measured at the same setup and compared with the values of the total reaction cross sections ${\sigma _{\rm{R}}}$ obtained in a number of other studies.

Each measurement session for a certain energy of the beam was conducted both with the target and without the target. The irradiation time was chosen in such a way that the number $I_0$ of events in both cases was approximately the same. The experimental information from all the detectors was recorded on a storage disk for subsequent offline analysis of each event of flight of a beam particle through the start $\Delta {E_0}$ detector, regardless of the reaction in the $\Delta {E_{\rm{T}}}$ detector. To reduce the effect of possible superposition of pulses in $\Delta {E_0}$ and $\Delta {E_{\rm{T}}}$ detectors, the beam intensity was limited to 10$^3$~s$^{-1}$.

\begin{figure}[htbp]
\includegraphics[width=8.6cm]{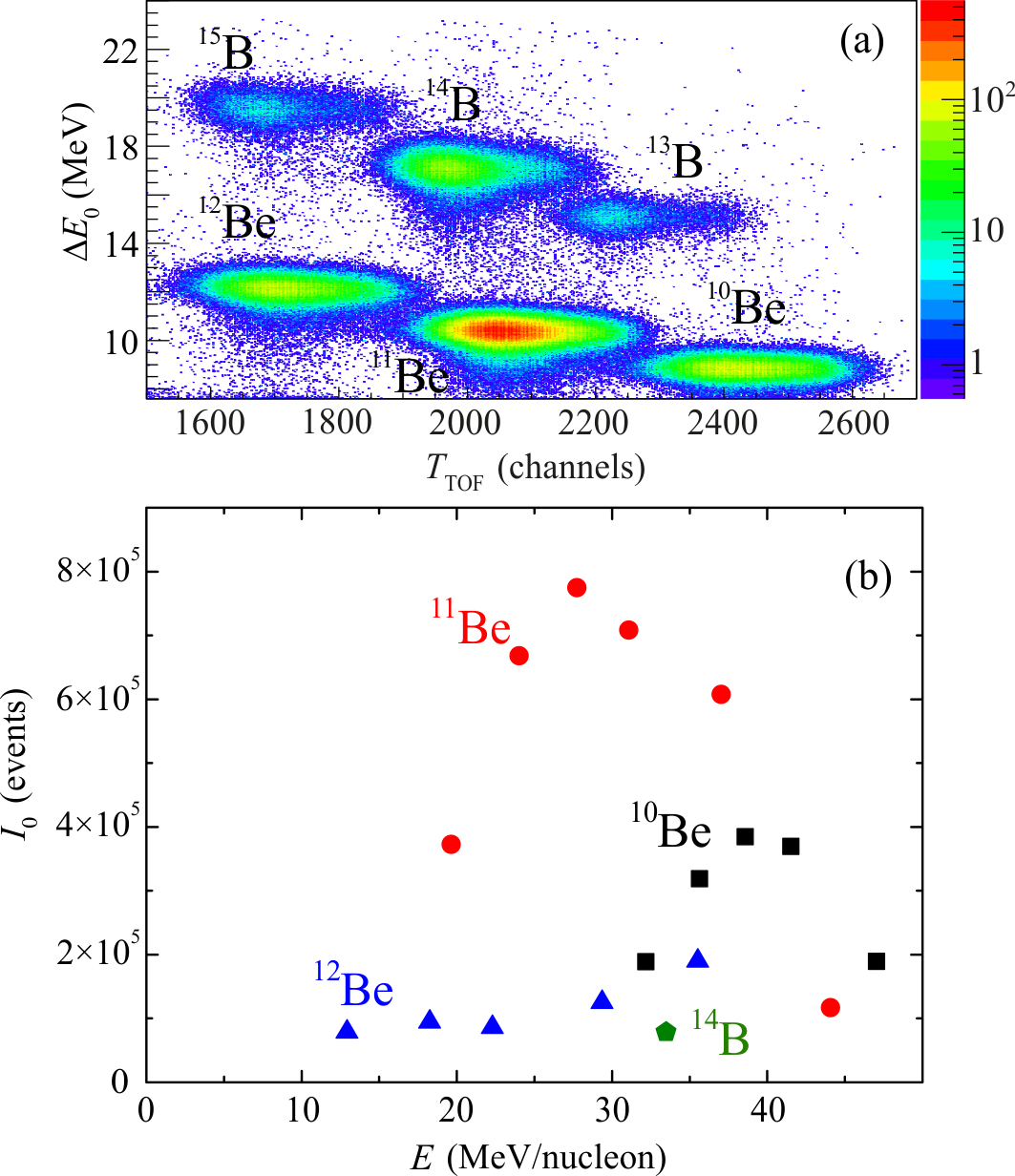}
\caption{\label{fig:fg03} (a) An example of a two-dimensional $\Delta {E_{\rm{0}}}$~${T_{{\rm{TOF}}}}$ matrix for of identification secondary beam particles $^{13,14,15}$B and $^{10,11,12}$Be nuclei in front of the target. The abscissa axis is the time of flight (inverted channels); the ordinate axis is the energy losses in the 243-$\mu$m thick $\Delta {E_{\rm{0}}}$ detector. (b) The total number of events ${I_{0}}$ for the $^{10,11,12}$Be and $^{14}$B nuclei: squares, circles, triangles, and pentagon respectively.}
\end{figure}

Particles of the secondary beam incident on the target were identified using two-dimensional $\Delta {E_{\rm{0}}}$~${T_{{\rm{TOF}}}}$ matrices, one of which is shown in Fig.~\ref{fig:fg03}(a). Each event of the passage of beam particles through the system of detectors located in front of the target is displayed by a point on the $\Delta {E_{\rm{0}}}$~${T_{{\rm{TOF}}}}$ matrix. Different colors correspond to different density of points according to the color scale gradation shown on the right. It can be seen that the points corresponding to the events of hitting the target by the $^{10,11,12}$Be and $^{14}$B nuclei fill well-separated regions on the $\Delta {E_{\rm{0}}}$~${T_{{\rm{TOF}}}}$ matrix corresponding to the mean energies of 47, 37, 29.4, and 33.5$A$~MeV, respectively. The condition of location of a point inside the given boundaries of a certain area on the two-dimensional $\Delta {E_{\rm{0}}}$~${T_{{\rm{TOF}}}}$ matrix allows one to reliably select a group of certain nuclei hitting the target, calculate their number $I_0$, and perform a subsequent analysis of the detected events $I_0$. The statistics for the $^{13}$B and $^{15}$B nuclei were considerably lower than those for other nuclei, therefore we did not include these nuclei in our analysis. The total number of events ${I_{0}}$ for the $^{10,11,12}$Be and $^{14}$B nuclei is shown in Fig.~\ref{fig:fg03}(b).

\begin{figure}[htbp]
\includegraphics[width=8.6cm]{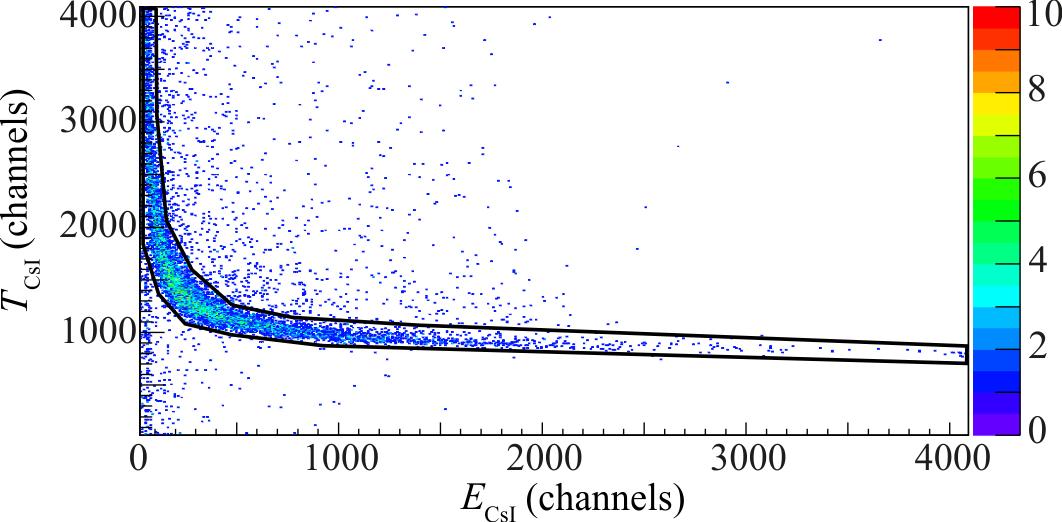}
\caption{\label{fig:fg04} An example of a two-dimensional ${E_{{\rm{CsI}}}}$~${T_{{\rm{CsI}}}}$ matrix from one of the 12 CsI(Tl) detectors of the spectrometer; ${E_{{\rm{CsI}}}}$ is the energy loss of the particles registered by the CsI(Tl) detector, ${T_{{\rm{CsI}}}}$ is the time interval between the pulses of the start detector and the CsI(Tl) detector.}
\end{figure}

First, the number $\Delta I$ of reaction events for $I_0$ was determined. Second, the distribution of the number of triggered detectors in $\Delta I$ reaction events was found. If at least one gamma quantum or neutron was detected in any of the twelve detectors of the spectrometer simultaneously with the detection of a particle that hit the target, such an event was considered a reaction event, and the number $\Delta I$ of such reaction events from the number ${I_0}$ of pre-selected events was determined. The condition of detection of a gamma quantum or a neutron in each CsI(Tl) detector was described by a contour on a two-dimensional amplitude-time ${E_{{\rm{CsI}}}}$~${T_{{\rm{CsI}}}}$ spectrum (matrix), where ${E_{{\rm{CsI}}}}$ is the energy loss of particles registered by the CsI(Tl) detector and ${T_{{\rm{CsI}}}}$ is the time interval between the pulses of the start detector and the CsI(Tl) detector of the spectrometer~\cite{BRAS_2019}. An example of the two-dimensional ${E_{{\rm{CsI}}}}$~${T_{{\rm{CsI}}}}$ matrix from one of the 12 CsI(Tl) detectors of the spectrometer is shown in Fig.~\ref{fig:fg04}. Each event of registration of a gamma quantum or neutron by the spectrometer detector is displayed on the two-dimensional ${E_{{\rm{CsI}}}}$~${T_{{\rm{CsI}}}}$ matrix. Different colors correspond to different density of points according to the color scale gradation shown on the right. The points corresponding to the registration of background particles are randomly distributed in the matrix, while the points corresponding to the registration of gamma quanta or neutrons from reactions in the target fill certain regions on the two-dimensional ${E_{{\rm{CsI}}}}$~${T_{{\rm{CsI}}}}$ matrix. Thus, the condition of the location of the point inside the region bounded by a contour on the $i$-th two-dimensional ${E_{{\rm{CsI}}}}$~${T_{{\rm{CsI}}}}$ matrix will determine the fact of registration of a $\gamma$-quantum or neutron by the $i$-th detector ($i = 1,...,{\rm{12}}$). The contour shape was determined in calibration measurements with a $^{60}$Co source. The boundaries of the regions were determined by the energy thresholds and time intervals for registration.

In experiments with thin targets, when the condition $\Delta I = {I_0} - I < < {I_0}$ is satisfied, formula~(\ref{eq:eq04}) may be reduced to the form
\begin{equation}
{\sigma _{\rm{R}}} = \frac{{\Delta I}}{{{I_0}n}}.
\label{eq:eq08}
\end{equation}
Ideally, a decrease $\Delta I$ of the number of particles due to inelastic reaction channels can be measured by a $4\pi $ detector covering the total solid angle around the target. In practice, the solid angle is smaller, and a quantity $\Delta \tilde I = \eta \Delta I$ is measured, where $\eta $ is the correction. The value $\Delta \tilde I$ was determined taking into account the detection efficiency of gamma quanta and neutrons by detectors.

\section{\label{sec:sec04} Detection efficiency of gamma quanta with different multiplicities}
In order to determine the detection efficiency $\varepsilon$ of gamma quanta with different multiplicities $M$ by the CsI(Tl) spectrometer (spectrometer response), measurements were performed with a $^{60}$Co calibration source installed instead of the target, a beta detector based on a thin plastic scintillator, and a trigger CeBr$_3$ scintillation detector with a PMT. These detectors were installed outside of the solid angle of the spectrometer. The beta detector, working in coincidence with the trigger detector, helped eliminate the background glow of the cerium bromide crystal caused by impurities.

The condition of $\beta$-$\gamma$ coincidences allowed us to single out the events of emission of one of the $\gamma$-quanta of the $^{60}$Co source which were registered by the CsI(Tl) detectors of the spectrometer. The half-life of the $^{60}$Co nucleus is convenient enough to use it as a source of gamma quanta. In the overwhelming majority (99.88\%) of the events of $\beta$-decay of $^{60}$Co, the daughter nucleus $^{60}$Ni$^{*}$ is formed in the excited state (4$^{+}$, 2.505~MeV) which decays with the emission of the $\gamma$-quantum with energy of ${E_{\gamma ,1}}$ = 1173~keV into the state $^{60}$Ni$^{*}$ (2$^{+}$, 1.332~MeV) followed by decay into the ground state of $^{60}$Ni (0$^{+}$, g.s.) with the emission of the $\gamma$-quantum with energy of ${E_{\gamma ,2}}$ = 1332~keV.

A set ${G_1}$ of events of registration of particles with energy release ${E_{\gamma ,2}}= 1332~\pm $~100~keV in the CeBr$_3$ scintillation detector was recorded. These events, in the overwhelming number of cases, correspond to the emission of gamma quantum with energy ${E_{\gamma ,1}}$ from $^{60}$Ni. The exceptions are background events that form a pedestal in the energy spectrum under the total absorption peak. The total number of events in the set ${G_1}$ is denoted by ${n_{1}}$. Taking into account averaging over the angles of emission of the gamma quantum ${E_{\gamma ,2}}$ entering the CeBr$_3$ detector, in the first approximation, the emission of the gamma quantum ${E_{\gamma ,1}}$ can be considered isotropic. Events from the set ${G_1}$ correspond to the processes of isotropic emission of a single gamma quantum with energy ${E_{\gamma ,1}}$ from the position of the target.

The registration and accumulation of the set ${G_1}$ of the events with multiplicity $M = 1$ and subsequent combining of records into the groups of two, three, etc., allowed us to obtain records of simulated events of $\gamma$-quantum emission with multiplicities $M = 2, 3,$ etc. These simulated events correspond to simultaneous emission from a source of two, three, etc. $\gamma$-quanta with energy of $E_{\gamma ,1} = 1173$~keV. We denote the number of the simulated events of emission of $M$ $\gamma$-quanta, in which $k$ spectrometer detectors were triggered, as $N_k^{(M)}$, where $k = 0, \ldots ,12$. The efficiency $\varepsilon _{{\rm{}}}(M)$ for registration of emission events with multiplicity $M$ by the spectrometer can be determined as the ratio of the total number of detected events with $k$ triggered detectors to the total number $n_M$ of events:
\begin{equation}
\varepsilon _{{\rm{}}}(M) = \frac{1}{{n_M }}\sum\limits_{k = 0}^{12} {N_k^{(M)} = \sum\limits_{k = 0}^{12} {w_M (k)}},
\label{eq:eq09}
\end{equation}
where
\begin{equation}
w_M (k) = \frac{1}{{n_M }}N_k^{(M)}
\label{eq:eq10}
\end{equation}
is the probability of triggering of $k$ detectors of the spectrometer when registering a cascade of gamma quanta with the multiplicity $M$.

\begin{figure}[htbp]
\includegraphics[width=6.6cm]{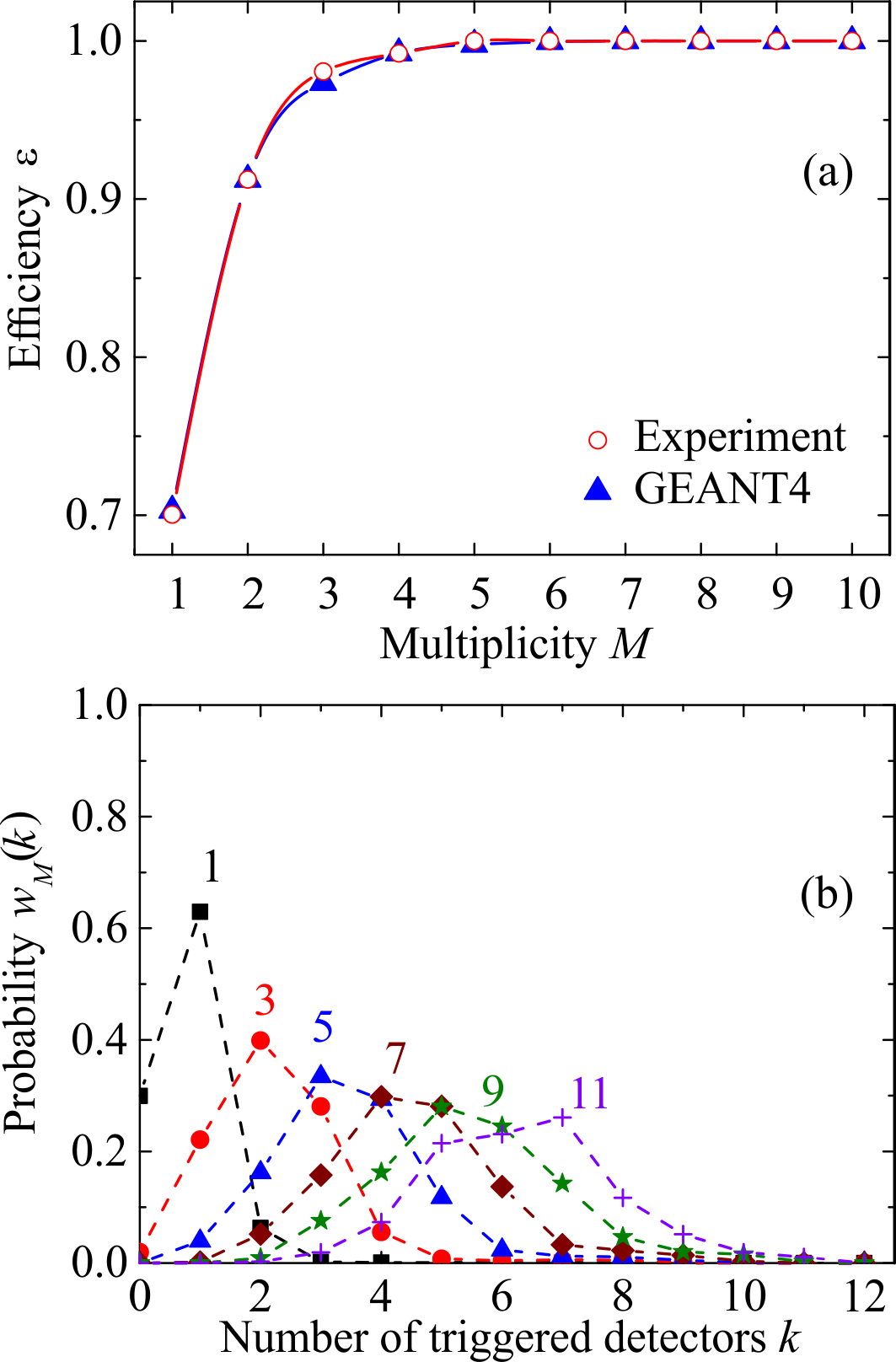}
\caption{\label{fig:fg05} (a) Experimental detection efficiency $\varepsilon _{{\rm{}}}(M)$ of $\gamma$-cascades with the multiplicity $M$ measured with the $^{60}$Co source for the 12-detector spectrometer (circles) and the results of the GEANT4 modeling (triangles). (b) Probabilities $w_M (k)$ of triggering of $k$ spectrometer detectors for experimentally simulated events of emission of $\gamma$-quanta with the multiplicity $M$; the values of $M$ are indicated.}
\end{figure}

The measured values of the detection efficiency $\varepsilon _{{\rm{}}}$ of simulated events with multiplicity $M$ for the $\gamma$-spectrometer with 12 detectors are shown in Fig.~\ref{fig:fg05}(a). Using the GEANT4 code~\cite{noauthor_geant4_nodate, geant} for this setup, we modeled the events of registration of the isotropic radiation of $\gamma$-quantum cascades with energy $E = 1173$~keV emitted from the center of the target into the total solid angle and determined the modeled detection efficiency which is also shown in Fig.~\ref{fig:fg05}(a). It can be seen that the modeled detection efficiency is in good agreement with the experimental data. This confirms the equivalence of combining the events of registration of single $\gamma$-quanta and measuring the events of the simultaneous emission of several $\gamma$-quanta. Therefore, the results obtained using the ${}^{60}$Co source were used to analyze the experimental data, when studying total reaction cross sections. The probabilities $w_M (k)$ of triggering of $k$ spectrometer detectors when registering simulated events of emission
of $\gamma$-cascades with multiplicity $M$ are shown in Fig.~\ref{fig:fg05}(b).

\section{\label{sec:sec05} Processing of experimental data}
Each measurement session at a certain beam energy was carried out both with and without a target. The sessions without the target were used to measure the background. Let ${N'_k}$ be the number of triggering of $k$ detectors for measurements without the target. The relationship between the number ${N'_k}$ and the number of particles ${I_0}$ can be approximated by a linear dependence
\begin{equation}
{N'_k} = {N'_{0\,k}} + {\beta _k}{I_0}.
\label{eq:eq11}
\end{equation}
The values of the coefficients ${\beta _k}$ and their uncertainties $\delta {\beta _k}$ can be determined using linear regression. In practice, the values of the parameters ${N'_{0\,k}}$ are small in magnitude, so we will use a simpler expression
\begin{equation}
{N'_{\,k}} = {\beta _k}{I_0},
\label{eq:eq12}
\end{equation}
with the coefficients ${\beta _k}$ found by the least squares method from the results of $m$ measurements
\begin{equation}
{\rm{ }}{\beta _k} = \frac{{\sum\limits_{j = 1}^m {{I_{0j}}{{N'}_{k\,j}}} }}{{\sum\limits_{j = 1}^m {I_{0j}^2} }}.
\label{eq:eq13}
\end{equation}
The numbers ${N'_k}$ of triggering of $k$ detectors for several energies $E$ and numbers ${I_0}$ of the $^{10,11,12}$Be and $^{14}$B nuclei with exposures without the target are given in Table~\ref{tab:tab02}. The coefficients ${\beta _k}$ and their uncertainties $\delta {\beta _k}$ are given in Tables~\ref{tab:tab03} and~\ref{tab:tab04}, respectively.
\begin{table*}[htbp]
\caption{\label{tab:tab02} The numbers ${N'_k}$ of triggering of $k$ detectors for several energies $E$ and numbers ${I_0}$ of the $^{10,11,12}$Be nuclei with exposures without the target.}
\begin{ruledtabular}
\begin{tabular}{ccccccccccccccc}
Nucleus & $E$ ($A$~MeV) & ${I_0}$ & ${N'_1}$ & ${N'_2}$ & ${N'_3}$ & ${N'_4}$ & ${N'_5}$ & ${N'_6}$ & ${N'_7}$ & ${N'_8}$ & ${N'_9}$ & ${N'_{10}}$ & ${N'_{11}}$ & ${N'_{12}}$\\
\colrule
$^{10}$Be & 35.6 & 188485 & 107 & 25 & 10 & 6 & 4 & 0 & 0 & 0 & 0 & 0 & 0 & 0\\
$^{11}$Be & 37.0 & 1471729 & 907 & 293 & 141 & 83 & 43 & 22 & 9 & 4 & 1 & 0 & 0 & 0\\
$^{12}$Be & 12.9 & 19909 & 19 & 13 & 2 & 0 & 0 & 0 & 0 & 0 & 0 & 0 & 0 & 0\\
$^{12}$Be & 22.3 & 47184 & 68 & 27 & 9 & 2 & 0 & 0 & 1 & 0 & 1 & 0 & 0 & 0\\
$^{12}$Be & 35.5 & 66750 & 107 & 53 & 9 & 6 & 2 & 2 & 1 & 0 & 0 & 0 & 0 & 0\\
$^{14}$B & 33.5 & 175156 & 129 & 26 & 17 & 1 & 2 & 2 & 3 & 1 & 0 & 0 & 0 & 0\\
\end{tabular}
\end{ruledtabular}
\end{table*}
\begin{table*}[htbp]
\caption{\label{tab:tab03} The values of the coefficients ${\beta _k}$ determined using the values of the parameters ${N'_{k}}$, ${I_0}$, and Eqs.~(\ref{eq:eq12}), (\ref{eq:eq13}).}
\begin{ruledtabular}
\begin{tabular}{cccccccccc}
Nucleus & $\beta_1$ & $\beta_2$ & $\beta_3$ & $\beta_4$ & $\beta_5$ & $\beta_6$ & $\beta_7$ & $\beta_8$ & $\beta_9$\\
\colrule
$^{10}$Be & $5.25 \times 10^{ - 4}$ & $1.47 \times 10^{ - 4}$ & $6.37 \times 10^{ - 5}$ & $3.74 \times 10^{ - 5}$ & $2.42 \times 10^{ - 5}$ & $7.86 \times 10^{ - 6}$ & $5.39 \times 10^{ - 6}$ & $3.62 \times 10^{ - 6}$ & $1.01 \times 10^{ - 6}$\\
$^{11}$Be & $5.89 \times 10^{ - 4}$ & $1.72 \times 10^{ - 4}$ & $8.70 \times 10^{ - 5}$ & $4.87 \times 10^{ - 5}$ & $2.70 \times 10^{ - 5}$ & $1.27 \times 10^{ - 5}$ & $5.33 \times 10^{ - 6}$ & $2.33 \times 10^{ - 6}$ & $9.04 \times 10^{ - 7}$\\
$^{12}$Be & $1.30 \times 10^{ - 3}$ & $5.87 \times 10^{ - 4}$ & $1.54 \times 10^{ - 4}$ & $4.16 \times 10^{ - 5}$ & $3.67\times 10^{ - 5}$ & $9.56 \times 10^{ - 6}$ & $4.95 \times 10^{ - 6}$ & $3.39 \times 10^{ - 6}$ & 0\\
$^{14}$B & $7.27 \times 10^{ - 4}$ & $1.45 \times 10^{ - 4}$ & $9.38 \times 10^{ - 5}$ & $6.46\times 10^{ - 6}$ & $1.09 \times 10^{ - 5}$ & $1.63 \times 10^{ - 5}$ & $5.45 \times 10^{ - 6}$ & 0 & 0\\
\end{tabular}
\end{ruledtabular}
\end{table*}
\begin{table*}[htbp]
\caption{\label{tab:tab04} The uncertainties $\delta {\beta _k}$ of the coefficients ${\beta _k}$ from Table~\ref{tab:tab03} determined using linear regression. For $^{14}$B, the values of $\delta {\beta _k}$ for $k=1,...,7$ are taken as the average values of $\delta {\beta _k}$ for the $^{10,11,12}$Be nuclei.}
\begin{ruledtabular}
\begin{tabular}{cccccccccc}
Nucleus & $ \delta \beta_1$ & $ \delta \beta_2$ & $ \delta \beta_3$ & $ \delta \beta_4$ & $ \delta \beta_5$ & $ \delta \beta_6$ & $ \delta \beta_7$ & $ \delta \beta_8$ & $ \delta \beta_9$\\
\colrule
$^{10}$Be & $6.02 \times 10^{ - 5}$ & $1.35 \times 10^{ - 5}$ & $1.75 \times 10^{ - 5}$ & $1.18 \times 10^{ - 5}$ & $7.23 \times 10^{ - 6}$ & $3.24 \times 10^{ - 6}$ & $2.46 \times 10^{ - 6}$ & $4.89 \times 10^{ - 6}$ & $1.01 \times 10^{ - 6}$\\
$^{11}$Be & $4.20 \times 10^{ - 5}$ & $2.86 \times 10^{ - 5}$ & $1.62 \times 10^{ - 5}$ & $8.32 \times 10^{ - 6}$ & $8.42 \times 10^{ - 6}$ & $3.49 \times 10^{ - 6}$ & $2.40 \times 10^{ - 6}$ & $2.86 \times 10^{ - 7}$ & $5.62 \times 10^{ - 7}$\\
$^{12}$Be & $6.71 \times 10^{ - 5}$ & $3.92 \times 10^{ - 5}$ & $7.81 \times 10^{ - 6}$ & $1.22 \times 10^{ - 5}$ & $4.64\times 10^{ - 6}$ & $3.39 \times 10^{ - 6}$ & $3.0 \times 10^{ - 6}$ & $6.26 \times 10^{ - 7}$ & 0\\
$^{14}$B & $5.64 \times 10^{ - 5}$ & $2.71 \times 10^{ - 5}$ & $1.38 \times 10^{ - 5}$ & $1.08 \times 10^{ - 5}$ & $6.76 \times 10^{ - 6}$ & $3.80 \times 10^{ - 6}$ & $2.62 \times 10^{ - 6}$ & 0 & 0\\
\end{tabular}
\end{ruledtabular}
\end{table*}

The numbers ${N_k}$ of triggering of $k$ detectors for several energies $E$ and numbers ${I_0}$ of the $^{10,11,12}$Be and $^{14}$B nuclei in exposures with the target are given in Table~\ref{tab:tab05}.
\begin{table*}[htbp]
\caption{\label{tab:tab05} The numbers ${N_k}$ of triggering of $k$ detectors for several energies $E$ and numbers ${I_0}$ of the $^{10,11,12}$Be and $^{14}$B nuclei in exposures with the target.}
\begin{ruledtabular}
\begin{tabular}{ccccccccccccccc}
Nucleus & $E$ ($A$~MeV) & ${I_0}$ & ${N_1}$ & ${N_2}$ & ${N_3}$ & ${N_4}$ & ${N_5}$ & ${N_6}$ & ${N_7}$ & ${N_8}$ & ${N_9}$ & ${N_{10}}$ & ${N_{11}}$ & ${N_{12}}$\\
\colrule
$^{10}$Be & 35.6 & 318920 & 597 & 436 & 345 & 258 & 173 & 86 & 43 & 20 & 8 & 5 & 0 & 0\\
$^{11}$Be & 37.0 & 607893 & 1279 & 1003 & 802 & 606 & 403 & 230 & 122 & 53 & 19 & 9 & 2 & 0\\
$^{12}$Be & 12.9 & 78713 & 246 & 198 & 120 & 89 & 47 & 32 & 20 & 12 & 1 & 0 & 0 & 0\\
$^{12}$Be & 22.3 & 85983 & 269 & 185 & 123 & 85 & 54 & 36 & 17 & 7 & 5 & 1 & 0 & 0\\
$^{12}$Be & 35.5 & 189834 & 543 & 385 & 245 & 185 & 143 & 62 & 46 & 21 & 10 & 0 & 0 & 0\\
$^{14}$B & 33.5 & 78573 & 168 & 144 & 124 & 95 & 65 & 37 & 25 & 14 & 7 & 0 & 0 & 0\\
\end{tabular}
\end{ruledtabular}
\end{table*}

Signals from gamma quanta and neutrons were not distinguished in the experiments. The neutron detection efficiency for the CsI(Tl) detectors used in the experiments was essentially lower than the gamma quanta detection efficiency. In the events of reactions accompanied by neutron emission, gamma quanta are also emitted, with the exception of elastic breakup channels~\cite{NIM_A274_1989}. For the above reasons, the values of ${w_M}(k)$ obtained by registering gamma quanta (Section~\ref{sec:sec04}) were used in the processing of the experimental data.

The results of measuring the total reaction cross section were determined taking into account the numbers of triggered detectors in the following order. Let $M$ photons and/or neutrons be emitted in a reaction with a probability $\Gamma (M)$, then the probability $P(k)$ of registration of the reaction event with triggering of $k$ detectors is
\begin{equation}
P(k)=\sum\limits_{M = 1}^{M_{\rm{max}}} {\Gamma (M){w_M}(k)}.
\label{eq:eq14}
\end{equation}
For the total number of registered reaction events $\Delta \tilde I = {I_0}{\tilde \sigma _{\rm{R}}}n$, the estimated number of their registrations with triggering of $k$ detectors is equal to
\begin{eqnarray}
P(k)\Delta \tilde I = {I_0}{\tilde \sigma_{\rm{R}}} n\sum\limits_{M = 1}^{M_{\rm{max}}}{\Gamma (M){w_M}(k)}.
\label{eq:eq15}
\end{eqnarray}
From the condition of equality of the number of registered events ${N_k} - {N'_k} = {N_k} - {\beta _k}{I_0}$ to their calculated value
\begin{eqnarray}
&&{N_k} - {\beta _k}{I_0} = {I_0} {\tilde \sigma _{\rm{R}}}n\sum\limits_{M = 1}^{M_{\rm{max}}} {\Gamma (M){w_M}(k)}\nonumber\\
&& = {I_0} n\sum\limits_{M = 1}^{M_{\rm{max}}} {{\tilde \sigma _{{\rm{R}}\,M}}{w_M}(k)},
\label{eq:eq16}
\end{eqnarray}
follows a system of linear equations
\begin{equation}
\sum\limits_{M = 1}^{M_{\rm{max}}} {{{\tilde \sigma }_{{\rm{R}}{\kern 1pt} M}}{w_M}(k)} - \frac{{{N_k} - {\beta _k}{I_0}}}{{{I_0}n}} = 0,
\label{eq:eq17}
\end{equation}
for the unknowns ${\tilde \sigma _{{\rm{R}}\,M}} = {\tilde \sigma _{\rm{R}}}\Gamma (M)$, $M\le M_{\rm{max}}$.

The coefficients of system~(\ref{eq:eq17}) are determined with uncertainties, therefore its exact solution may yield non-physical values ${\tilde \sigma _{{\rm{R}}\,M}} < 0$. Thus, it is more correct to find these values from the condition that the sum of the squares of the left-hand sides is minimum
\begin{eqnarray}
&&F\left( {{{\tilde \sigma }_{{\rm{R}}\,1}}, \ldots ,{{\tilde \sigma }_{{\rm{R}}\,{M_{\rm{max}}}}}} \right)\nonumber\\
&& = \sum\limits_{k = 1}^{12} {{{\left[ {\sum\limits_{M = 1}^{M_{\rm{max}}} {{{\tilde \sigma }_{{\rm{R}}\,M}}{w_M}(k) - \frac{{{N_{k}} - {\beta _k}{I_0}}}{{{I_0}n}}} } \right]}^2}},
\label{eq:eq18}
\end{eqnarray}
under the constraint ${\tilde \sigma _{{\rm{R}}{\kern 1pt} M}} \ge 0$.

\begin{figure}[htbp]
\includegraphics[width=7.6cm]{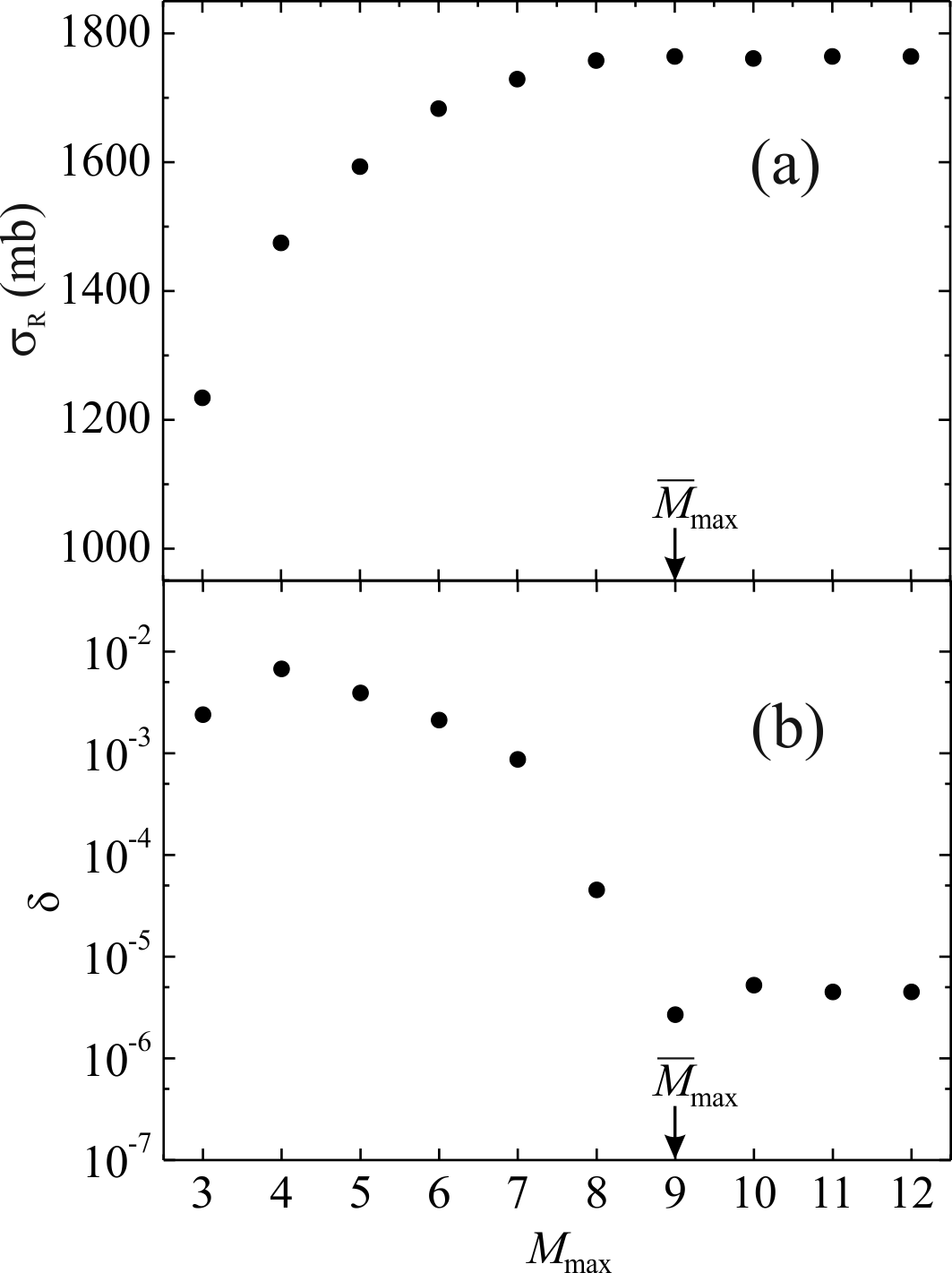}
\caption{\label{fig:fg06} (a) The raw value of the total cross section for the reaction $^{11}$Be~+~$^{28}$Si at an energy of 37$A$~MeV. (b) The minimum $\delta$ as function of $M_{\rm{max}}$ in formula~(\ref{eq:eq19}). The arrow indicates the absolute minimum of $\delta$ at ${{M}_{\max }} =9$.}
\end{figure}

The values of
\begin{equation}
{\tilde \sigma _{\rm{R}}}(M_{\rm{max}}) = \sum\limits_{M = 1}^{M_{\rm{max}}} {{{\tilde \sigma }_{{\rm{R}}\,M}}},
\label{eq:eq19}
\end{equation}
as a function of $M_{\rm{max}}$ are shown in Fig.~\ref{fig:fg06}(a) for the reaction $^{11}$Be~+~$^{28}$Si at an energy of 37$A$~MeV. We see that convergence takes place for $M_{\rm{max}} > 8$. The minimum
\begin{equation}
\delta(M_{\rm{max}}) = F\left( {{{\tilde \sigma }_{{\rm{R}}\,1}}, \ldots ,{{\tilde \sigma }_{{\rm{R}}\,{M_{\rm{max}}}}}} \right)
\label{eq:eq20}
\end{equation}
is shown in Fig.~\ref{fig:fg06}(b). From the condition of the absolute minimum of the values of $\delta(M_{\rm{max}})$, we determined the values of ${{\bar M}_{\max }}$ for all reactions.

\begin{figure}[htbp]
\includegraphics[width=6.0cm]{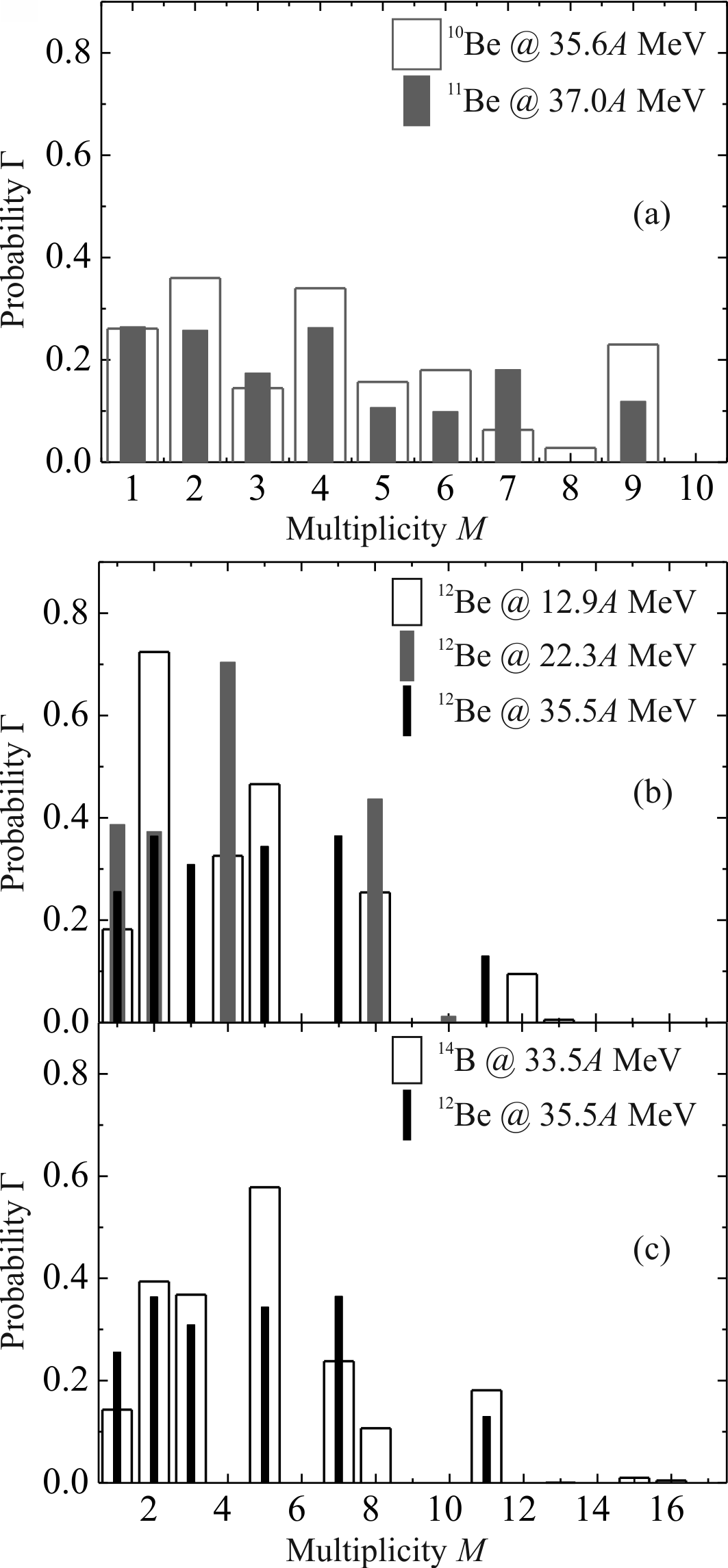}
\caption{\label{fig:fg07} The obtained probability $\Gamma (M)$ of the emission of $M$ photons and/or neutrons in the reactions for (a) $^{10}$Be at an energy of 35.6$A$~MeV (grey bars) and $^{11}$Be at 37.0$A$~MeV (empty bars); (b) $^{12}$Be at 12.9$A$~MeV (empty bars), 22.3$A$~MeV (grey bars), and 35.5$A$~MeV (black bars); (c) $^{14}$B at 33.5$A$~MeV (empty bars) and $^{12}$Be at 35.5$A$~MeV (black bars).}
\end{figure}

The examples of the obtained probability $\Gamma \left( M \right)$ of the emission of $M$ photons and/or neutrons in the studied reactions are shown in
Fig.~\ref{fig:fg07}.

The raw values of the total reaction cross sections were calculated as the sums
\begin{equation}
{\tilde \sigma _{\rm{R}}} = \sum\limits_{M = 1}^{{\bar M}_{\max }} {{{\tilde \sigma }_{{\rm{R}}\,M}}}.
\label{eq:eq21}
\end{equation}
The uncertainty in $\Delta {\tilde \sigma _{\rm{R}}}$ due to the uncertainty $\delta {\beta _k}$ of the coefficients ${\beta _k}$ was calculated by formula
\begin{equation}
\Delta {\tilde \sigma _{\rm{R}}} = {{\left| {\tilde \sigma _{\rm{R}}^{( + )} - \tilde \sigma _{\rm{R}}^{( - )}} \right|} \mathord{\left/ {\vphantom {{\left| {\tilde \sigma _{\rm{R}}^{( + )} - \tilde \sigma _{\rm{R}}^{( - )}} \right|} 2}} \right. \kern-\nulldelimiterspace} 2},
\label{eq:eq23}
\end{equation}
where ${\tilde \sigma _{\rm{R}}^{( + )}}$ and ${\tilde \sigma _{\rm{R}}^{( - )}}$ are the values for the set of parameters ${\beta _k} + \delta {\beta _k}$ and ${\beta _k} - \delta {\beta _k}$, respectively.

The total reaction cross section $\sigma _{\rm{R}}$ was found as
\begin{equation}
{\sigma _{\rm{R}}} = {{{{\tilde \sigma }_{\rm{R}}}} \mathord{\left/ {\vphantom {{{{\tilde \sigma }_{\rm{R}}}} \eta }} \right. \kern-\nulldelimiterspace} \eta }.
\label{eq:eq22}
\end{equation}
The values of the correction $\eta$ and its errors $\Delta \eta$ were determined by the condition that the confidence intervals of the cross sections for the reactions $^{10,11,12}$Be~+~$^{28}$Si and $^{14}$B~+~$^{28}$Si coincide with the known data from other studies. Thus, the uncertainties of the correction $\eta$ are the result of the uncertainties in the experimental data.
The values of the relative ${\varepsilon _\sigma }$ and absolute $\Delta {\sigma _{\rm{R}}}$ uncertainties were calculated by formulas,
\begin{equation}
{\varepsilon _\sigma } = \frac{{\Delta {{\tilde \sigma }_{\rm{R}}}}}{{{{\tilde \sigma }_{\rm{R}}}}} + \frac{{\Delta \eta }}{\eta },
\label{eq:eq24}
\end{equation}
\begin{equation}
\Delta {\sigma _{\rm{R}}} = {\sigma _{\rm{R}}}{\varepsilon _\sigma }.
\label{eq:eq25}
\end{equation}

The uncertainties ${\delta _E}$ associated with the spread of the beam energy before the target were determined from the analysis of one-dimensional spectra $\Delta {E_{\rm{0}}}$ and ${T_{{\rm{TOF}}}}$. The energy losses ${E_1}$ of the projectile nuclei at the exit from the target were calculated using the \textsc{lise++} code~\cite{noauthor_lise++_nodate, lise}. The spread of energy losses in the target was determined by the expression ${\Delta _E} = {{\left( {{E_0} - {E_1}} \right)} \mathord{\left/ {\vphantom {{\left( {{E_0} - {E_1}} \right)} 2}} \right. \kern-\nulldelimiterspace} 2} > > {\delta _E}$. The value of the cross section was attributed to the value of energy ${E_0} - {\Delta _E}$.

\section{\label{sec:sec06} Experimental results and discussion}
\begin{figure}[htbp]
\includegraphics[width=7.6cm]{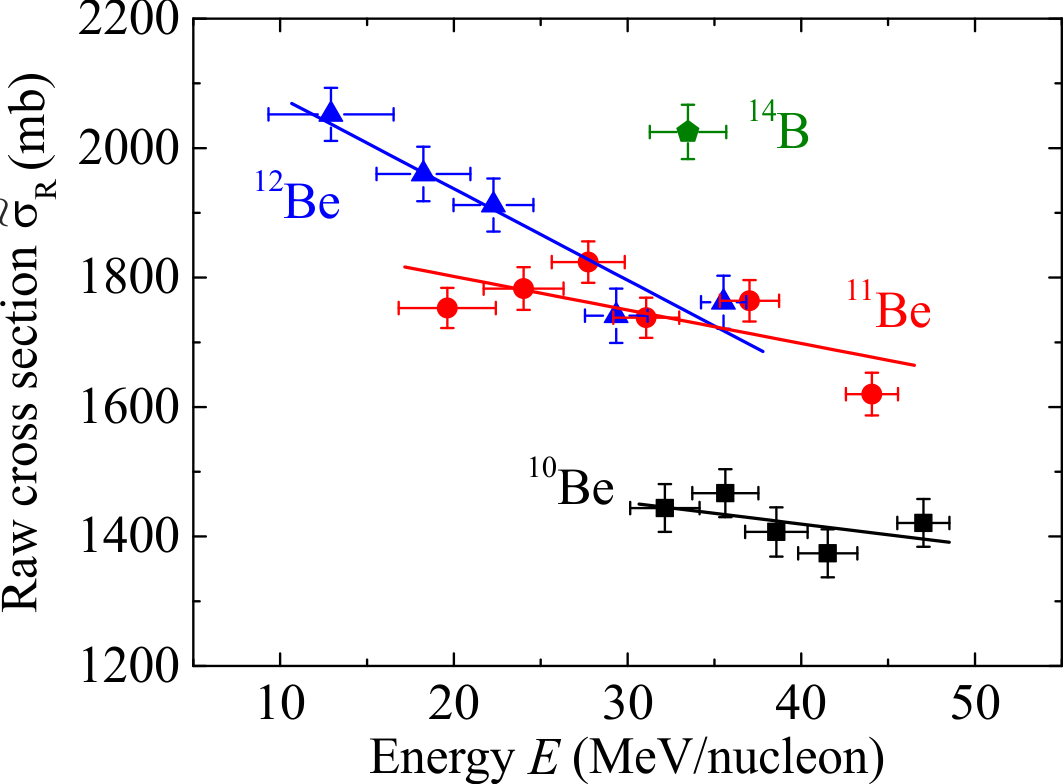}
\caption{\label{fig:fg08} Experimental total cross sections $\tilde \sigma _{\rm{R}}$ obtained by formula~(\ref{eq:eq21}) (raw data) for the reactions $^{10,11,12}$Be~+~$^{28}$Si (this work): $^{10}$Be (squares), $^{11}$Be (circles), $^{12}$Be (triangles), $^{14}$B (pentagon). Solid lines are the results of linear regression.}
\end{figure}

The experimental results on the total cross sections $\tilde \sigma _{\rm{R}}$ obtained by formula~(\ref{eq:eq21}) (raw data) for the reactions $^{10,11,12}$Be~+~$^{28}$Si and $^{14}$B~+~$^{28}$Si are shown in Fig.~\ref{fig:fg08} and given in Table~\ref{tab:tab06}.
\begin{table}[b]
\caption{\label{tab:tab06} Experimental values of the total reaction cross sections as a function of the collision energy $E$: raw data $\tilde \sigma _{\rm{R}}$ and ${\sigma _{\rm{R}}}$ taking into account correction $\eta$.}
\begin{ruledtabular}
\begin{tabular}{ccccc}
Nucleus & $E$ ($A$~MeV) & $\tilde \sigma _{\rm{R}}$ (mb) & $\eta$ & ${\sigma _{\rm{R}}}$ (mb)\\
\colrule
$^{10}$Be & 32.2 $\pm$ 2.0 & 1444 $\pm$ 37 & 0.92 $\pm$ 0.02 & 1587 $\pm$ 68\\
 & 35.6 $\pm$ 1.9 & 1467 $\pm$ 37 & & 1612 $\pm$ 67\\
 & 38.6 $\pm$ 1.8 & 1407 $\pm$ 38 & & 1549 $\pm$ 68\\
 & 41.6 $\pm$ 1.7 & 1374 $\pm$ 37 & & 1509 $\pm$ 67\\
 & 47.0 $\pm$ 1.5 & 1421 $\pm$ 37 & & 1562 $\pm$ 67\\
$^{11}$Be & 19.6 $\pm$ 2.8 & 1753 $\pm$ 31 & 0.78 $\pm$ 0.03 & 2247 $\pm$ 127\\
 & 24.0 $\pm$ 2.3 & 1783 $\pm$ 33 & & 2284 $\pm$ 129\\
 & 27.7 $\pm$ 2.1 & 1824 $\pm$ 32 & & 2338 $\pm$ 130\\
 & 31.1 $\pm$ 1.9 & 1738 $\pm$ 31 & & 2230 $\pm$ 127\\
 & 37.0 $\pm$ 1.7 & 1764 $\pm$ 32 & & 2261 $\pm$ 127\\
 & 44.1 $\pm$ 1.5 & 1620 $\pm$ 33 & & 2083 $\pm$ 122\\
$^{12}$Be & 12.9 $\pm$ 3.6 & 2052 $\pm$ 41 & 0.9 $\pm$ 0.03 & 2332 $\pm$ 124\\
 & 18.3 $\pm$ 2.7 & 1960 $\pm$ 42 & & 2227 $\pm$ 120\\
 & 22.3 $\pm$ 2.3 & 1912 $\pm$ 41 & & 2173 $\pm$ 118\\
 & 29.4 $\pm$ 1.8 & 1741 $\pm$ 42 & & 1978 $\pm$ 113\\
 & 35.5 $\pm$ 1.3 & 1768 $\pm$ 42 & & 2009 $\pm$ 115\\
$^{14}$B & 33.5 $\pm$ 2.2 & 2025 $\pm$ 35 & 0.84 $\pm$ 0.05 & 2411 $\pm$ 143\\
\end{tabular}
\end{ruledtabular}
\end{table}

The raw cross section for the reaction $^{10}$Be~+~$^{28}$Si has the smallest values among all reactions and slowly increases with decreasing energy of the $^{10}$Be beam. The raw cross sections for the $^{11,12}$Be nuclei exceed the raw cross section for the $^{10}$Be nucleus on average by $\sim$~600~mb in the energy range of 30$A$--50$A$~MeV. This fact may be explained as a result of a larger interaction radius of the $^{11,12}$Be nuclei. The raw cross section for the $^{11}$Be nucleus slowly depends on energy. The raw cross section for the $^{12}$Be nucleus sharply and almost linearly increases with decreasing energy in the energy range of 10$A$--35$A$~MeV. A similar energy dependence of the total reaction cross section in the same energy range was observed for the reaction $^{9}$Li~+~$^{28}$Si~\cite{penionzhkevich_peculiarities_2017}. This behavior may be explained by the skin structure of the outer neutron shell of the $^{9}$Li and $^{12}$Be nuclei. The differences in the raw cross sections for the reactions of $^{11}$Be and $^{10}$Be, as well as $^{14}$B and $^{12}$Be are similar, $\sim$~600~mb, which may be explained as a result of additional halo structures in the $^{11}$Be and $^{14}$B nuclei.

The results for the total reaction cross sections $ {\sigma }_{\rm{R}} (E)$ obtained with the correction $\eta$ are also given in Table~\ref{tab:tab06}. The differences between the raw cross sections and the cross sections with the correction $\eta$ for the reactions with $^{10}$Be are $\sim$~150~mb, for $^{12}$Be and $^{14}$B -- $\sim$~300~mb, and for $^{11}$Be -- $\sim$~500~mb. As mentioned above, this fact may be explained as a result of emission of outer neutrons with different neutron binding energies (Table~\ref{tab:tab01}) into small forward angles.

\begin{figure}[htbp]
\includegraphics[width=6.6cm]{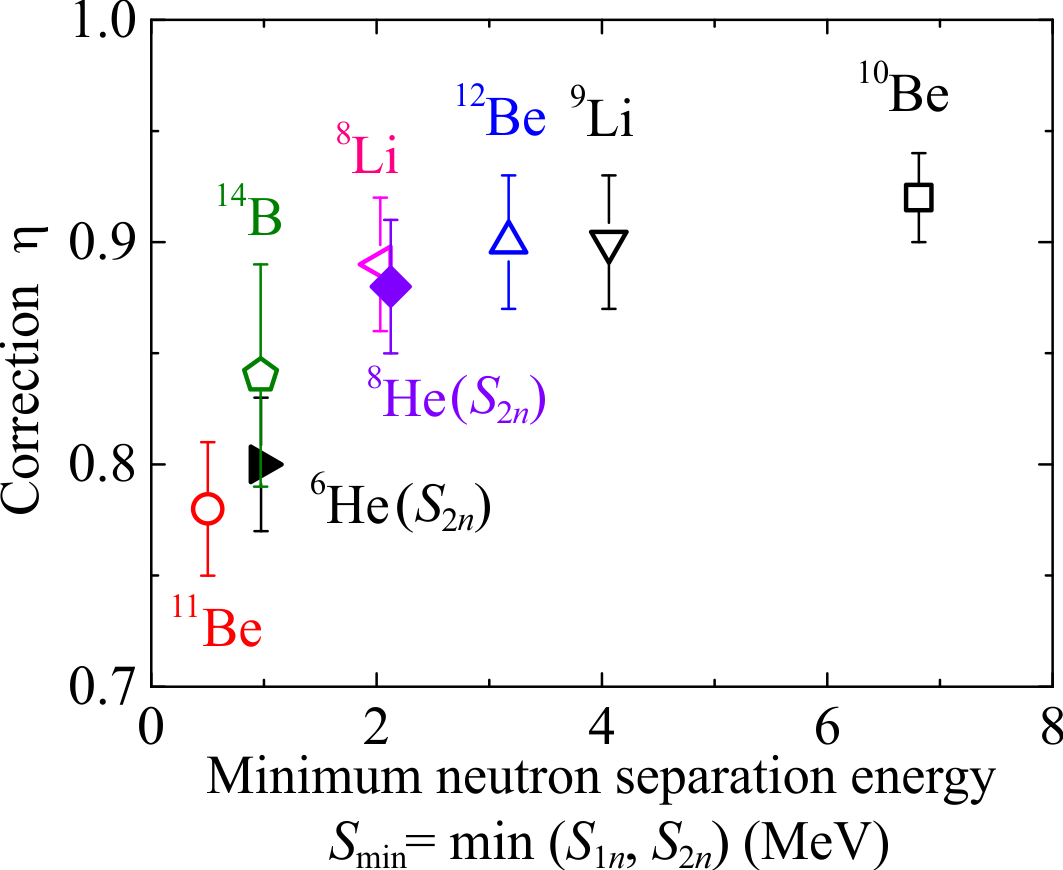}
\caption{\label{fig:fg09} The dependence of the correction $\eta$ on the value of the minimum neutron separation energy ${S_{\min }} = \min \left( {{S_{1n}},{S_{2n}}} \right)$ for the nuclei $^{10}$Be (square), $^{11}$Be (circle), $^{12}$Be (triangle up), $^{14}$B (pentagon), $^{6}$He (triangle right)~\cite{Sobolev_2020}, $^{8}$He (diamond)~\cite{Sobolev_2020}, $^{8}$Li (triangle left)~\cite{www}, $^{9}$Li (triangle down)~\cite{Sobolev_2020}. Empty symbols correspond to ${{S_{1n}}}$, filled symbols --- to ${{S_{2n}}}$.}
\end{figure}

The obtained dependence of the correction $\eta$ on the value of the minimum neutron separation energy ${S_{\min }} = \min \left( {{S_{1n}},{S_{2n}}} \right)$ for the nuclei $^{6,8}$He, $^{8,9}$Li, $^{10,11,12}$Be, and $^{14}$B is shown in Fig.~\ref{fig:fg09}. We see that the results for the isotopes of Be and B are consistent within the errors with the previously obtained systematic dependence of $\eta \left( {{S_{\min }}} \right)$ found for neutron-rich isotopes of He and Li~\cite{Sobolev_2020, www}. The value of the correction $\eta$ may be an indication of the degree of anisotropy of neutron emission: the smaller the values of $\eta$, the larger the fraction of neutrons emitted into the small forward angles. For the $^{A}$Be isotopes with $A>10$ having dumbbell-shaped core \{$^{10}$Be\}, the value of $\eta$ depends on the distribution of the outer neutrons on the levels in the shell model of the deformed nucleus.

\begin{figure}[htbp]
\includegraphics[width=7.6cm]{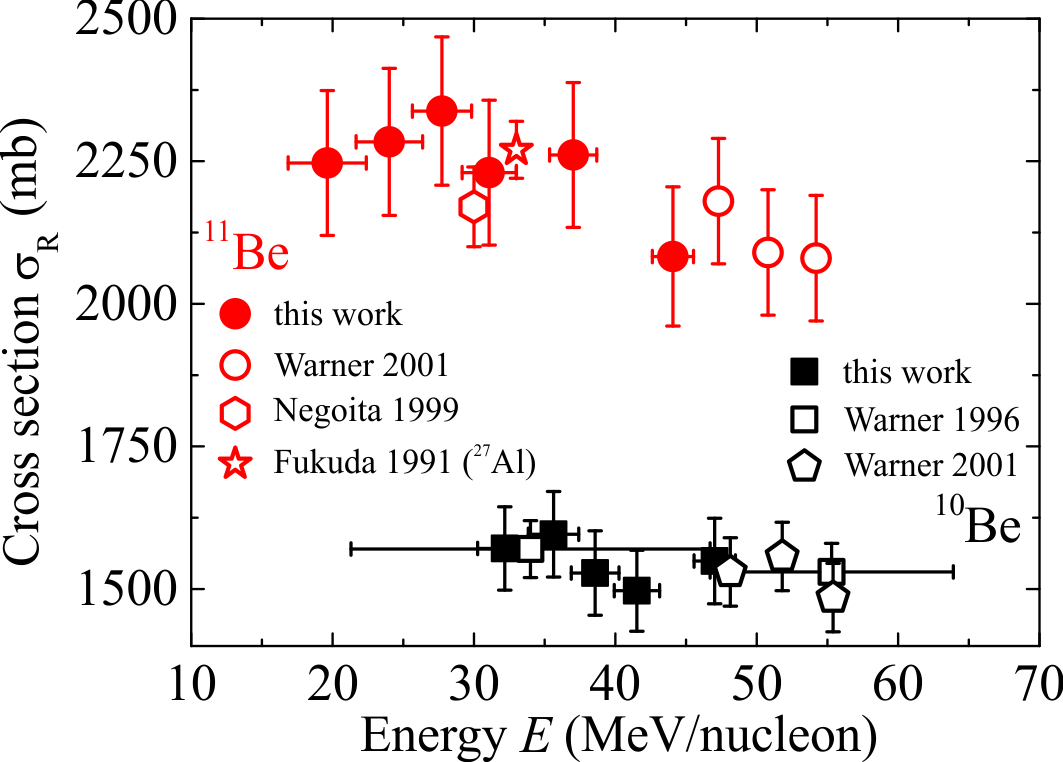}
\caption{\label{fig:fg10} Experimental total cross sections for the reactions $^{10}$Be~+~$^{28}$Si: filled squares (this work), empty squares (Warner 1996~\cite{warner_total_1996}), pentagons (Warner 2001~\cite{warner_2001}); $^{11}$Be~+~$^{28}$Si: filled circles (this work), empty circles (Warner 2001~\cite{warner_2001}), hexagon (Negoita 1999~\cite{Negoita}); $^{11}$Be~+~$^{27}$Al: star (Fukuda 1991~\cite{Fukuda_1991}).}
\end{figure}

The experimental total cross sections $ {\sigma }_{\rm{R}} (E)$ for the $^{10,11}$Be~+~$^{28}$Si reactions are presented in Fig.~\ref{fig:fg10} in comparison with the data from other studies. The cross section for the $^{11}$Be~+~$^{28}$Si reaction coincides within errors with the cross section for the $^{11}$Be~+~$^{27}$Al reaction. It is well known that the experimental signs of a neutron halo are an anomalously large total reaction cross section, a narrow momentum distribution of fragmentation or breakup products, and a low value the neutron separation energy for a nucleus~\cite{Hansen1987}. Therefore, the large difference between the total reaction cross sections for the $^{10}$Be and $^{11}$Be nuclei along with a low value of the neutron separation energy for $^{11}$Be (${S_{1n}}=$~0.5~MeV) is a clear indication of the halo structure of the $^{11}$Be nucleus, which is consistent with the conclusions of other works~\cite{Hansen1987, warner_2001}.

\begin{figure}[htbp]
\includegraphics[width=7.6cm]{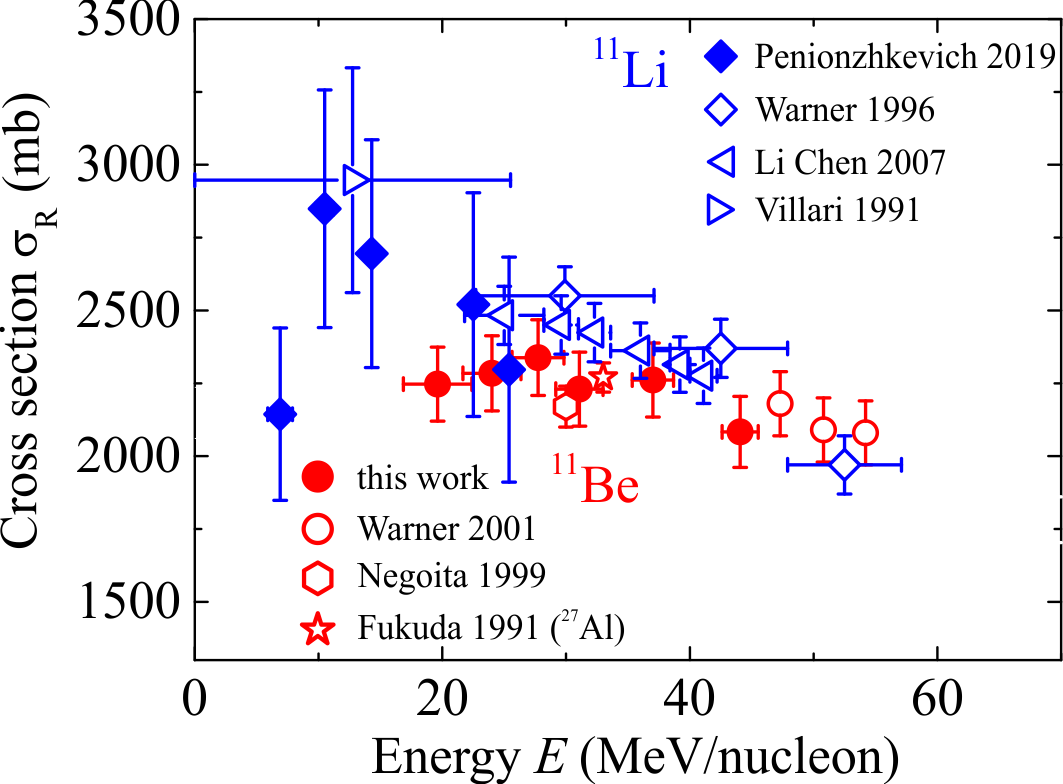}
\caption{\label{fig:fg11} Experimental total cross sections for the reactions $^{11}$Be~+~$^{28}$Si: filled circles (this work), empty circles (Warner 2001~\cite{warner_2001}), hexagon (Negoita 1999~\cite{Negoita}); $^{11}$Be~+~$^{27}$Al: star (Fukuda 1991~\cite{Fukuda_1991}); $^{11}$Li~+~$^{28}$Si: filled diamonds (Penionzhkevich 2019~\cite{Penionzhkevich_2019}), empty diamonds (Warner 1996~\cite{warner_total_1996}), triangles left (Li Chen 2007~\cite{li_chen_measurement_2007}), triangle right (Villari 1991~\cite{Villari}).}
\end{figure}

The experimental total cross sections $ {\sigma }_{\rm{R}} (E)$ for the $^{11}$Be~+~$^{28}$Si reaction are presented in Fig.~\ref{fig:fg11} in comparison with the data for the $^{11}$Li~+~$^{28}$Si reaction and the data from other studies. The cross section values for both reactions are similar, which is consistent with the similar and very small neutron separation energies ${S_{1n}}$ for the $^{11}$Be and $^{11}$Li nuclei, 0.5~MeV and 0.4~MeV, respectively. We see a sharp increase in the total reaction cross section for $^{11}$Li at energies $\sim$~10$A$~MeV; the cross section is anomalously large in comparison with $^{9}$Li~\cite{Penionzhkevich_2019}. Small differences in $ {\sigma }_{\rm{R}} (E)$ for the $^{11}$Be and $^{11}$Li nuclei may be due to the difference in their deformations ($^{11}$Li is oblate with the quadrupole deformation parameter $\beta_2=-0.636$, e.g.,~\cite{cdfe}).

Note that the probably oblate $^{10}$Li nucleus (as oblate, as $^{11}$Li) and the spherical $^{5}$He nucleus with one unpaired outer neutron are unbound, while the prolate $^{11}$Be nucleus with one unpaired outer neutron is bound. The effect may be due to the large prolate deformation of the latter. We may illustrate this phenomenon by a humorous example of the sea attraction --- riding an inflatable banana-shape boat. One can easily sit on the quickly moving banana-shape boat. However, no one even tries to ride a ball-shape boat. So this effect of the bound neutron state in the prolate $^{11}$Be nucleus and unbound neutron states in the probably oblate $^{10}$Li nucleus and spherical $^{5}$He nucleus may be called a banana effect.

\begin{figure}[htbp]
\includegraphics[width=7.6cm]{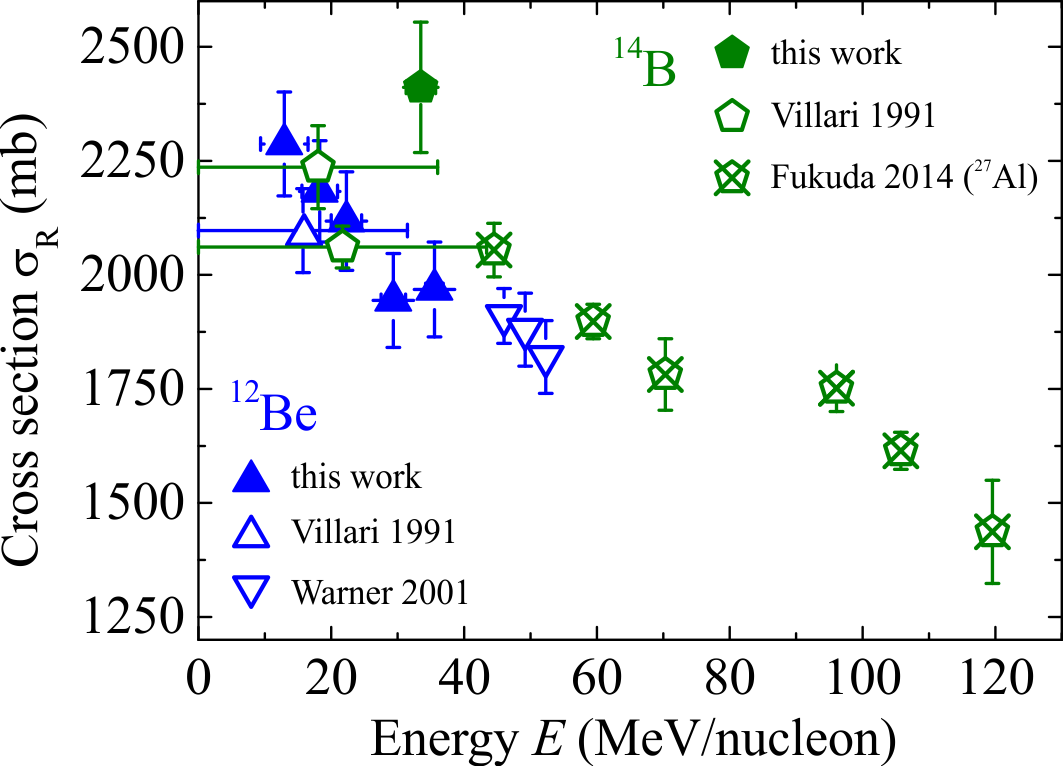}
\caption{\label{fig:fg12} Experimental total cross sections for the reactions $^{12}$Be~+~$^{28}$Si: filled triangles (this work), empty triangle up (Villari 1991~\cite{Villari}), empty triangles down (Warner 2001~\cite{warner_2001}); $^{14}$B~+~$^{28}$Si: filled pentagon (this work), empty pentagons (Villari 1991~\cite{Villari}); $^{14}$B~+~$^{27}$Al: crossed pentagons (Fukuda 2014~\cite{Fukuda_2014}).}
\end{figure}

The experimental total cross sections $ {\sigma }_{\rm{R}} (E)$ for the $^{12}$Be,$^{14}$B~+~$^{28}$Si reactions are presented in Fig.~\ref{fig:fg12} in comparison with the data from other studies. The total cross section for the $^{12}$Be~+~$^{28}$Si reaction is also compared with the total cross section for the $^{14}$B~+~$^{27}$Al reaction. For these reactions, we see the close values of the total cross sections and a similar increase with decreasing energy. An increase in the total reaction cross sections in the region below 20$A$~MeV can be explained by a change in the nucleon distribution when the nuclei approach each other, which leads to a decrease in the height of the Coulomb barrier, as in the reaction $^{9}$Li~+~$^{28}$Si~\cite{penionzhkevich_peculiarities_2017}.

From Figs.~\ref{fig:fg10}--\ref{fig:fg12}, we can see that the obtained experimental data are in agreement with the results of other studies and also cover the previously unexplored range of low energies.

\section{\label{sec:sec10} Conclusions}
We measured the total reaction cross sections for the $^{10,11,12}$Be and $^{14}$B nuclei on the $^{28}$Si target in the beam energy range $13A$--$47A$~MeV by the direct and model-independent 4$\pi$-method based on the registration of the $\gamma$-quanta and neutrons accompanying the interaction by the multi-detector spectrometer. The procedure of processing of the obtained experimental data was based on taking into account the number of triggered spectrometer detectors.

It was found that the values of the cross sections for $^{11}$Be and $^{12}$Be are significantly larger than those for $^{10}$Be. For the $^{11}$Be nucleus, the large values of the measured total reaction cross section and the low value of the neutron separation energy (0.5~MeV) indicate its halo structure. For the $^{12}$Be nucleus, the large values of the measured total reaction cross section, two paired outer neutrons, and the larger value of the neutron separation energy (3.2~MeV) indicate a nuclear structure, which is more compact than a halo and can be called a skin.

Therefore, the study of the total cross sections for the reactions involving neutron-rich weakly bound nuclei makes it possible to obtain information on their structure (halo, skin, clustering, etc.) and its manifestation in nuclear reactions.

\begin{acknowledgments}
The authors express their gratitude to A.~A.~Bezbakh and S.~A.~Krupko from the scientific group of the ACCULINNA fragment separator for all possible assistance in carrying out experiments. The authors thank the HybriLIT team for the technical support of calculations on the Heterogeneous Cluster of the Joint Institute for Nuclear Research.
\end{acknowledgments}


\end{document}